\documentclass[lettersize,journal]{IEEEtran}

\usepackage{amsfonts}
\usepackage{amssymb}
\usepackage{cite}
\usepackage[cmex10]{amsmath}
\usepackage{float}
\usepackage{color}
\usepackage{stfloats,fancyhdr}
\usepackage{amsmath}
\usepackage{amsthm}
\usepackage{enumerate}
\usepackage{array}
\usepackage{algpseudocode,algorithm}

\usepackage{multirow}
\usepackage{changepage}
\usepackage[normalem]{ulem}

\usepackage{url}
\usepackage{xr-hyper}
\usepackage{epstopdf}
\usepackage{tablefootnote}

\newtheorem{corollary}{Corollary}

\newtheorem{proposition}{\textbf{Proposition}}

\ifCLASSINFOpdf
\usepackage[pdftex]{graphicx}
\DeclareGraphicsExtensions{.pdf,.jpeg,.png}
\else
\usepackage[dvips]{graphicx}
\DeclareGraphicsExtensions{.eps}
\fi

\usepackage{subfigure}
\usepackage{fancybox,dashbox}

\usepackage{amsmath}

\usepackage{array}

\begin{document}
		%
		\title{A Just-In-Time Networking Framework for Minimizing Request-Response Latency of Wireless Time-Sensitive Applications}
		%
		%
		%
		%
		
		\author{Lihao~Zhang,
			Soung~Chang~Liew,~\IEEEmembership{Fellow,~IEEE,}
			and~He~Chen,~\IEEEmembership{Member,~IEEE}%
			\thanks{This work was supported in part by xxx. \textit{(Corresponding author: Soung Chang Liew).}}
			\thanks{Lihao Zhang, Soung Chang Liew and He Chen are with Department of Information Engineering, The Chinese University of Hong Kong, Hong Kong SAR, China (email: zl018@ie.cuhk.edu.hk; soung@ie.cuhk.edu.hk; he.chen@ie.cuhk.edu.hk).}
		}
		
		%
		%

	\markboth{Journal of \LaTeX\ Class Files,~Vol.~xx, No.~x, xx~xxxx}%
	{Shell \MakeLowercase{\textit{et al.}}: Bare Demo of IEEEtran.cls for Computer Society Journals}

\maketitle

		\begin{abstract}
			This paper puts forth a networking paradigm, referred to as just-in-time (JIT) communication, to support client-server applications with stringent request-response latency requirement. Of interest is not just the round-trip delay of the network, but the actual request-response latency experienced by the application. The JIT framework contains two salient features. At the client side, the communication layer will “pull” a request from the client just when there is an upcoming transmission opportunity from the network. This ensures that the request contains information that is as fresh as possible (e.g., a sensor reading obtained just before the transmission opportunity). At the server side, the network ascertains that the server, after receiving and processing the request to generate a response (e.g., a control command to be sent to the client), will have a transmission opportunity at just this time. We realize the JIT system, including the protocol stack, over a Time-Division-Multiple-Access (TDMA) network implemented on a System-on-Chip (SoC) platform. We prove that a TDMA network with a power-of-2 time slots per superframe is optimal for realizing the server-side JIT function. Our experimental results validate that JIT networks can yield significantly lower request-response latency than networks without JIT support can.
		\end{abstract}
		

	\section{Introduction}\label{sec:introduction}
	\IEEEPARstart{T}{he} emerging time-sensitive applications in automotive, avionics, building management, and industrial automation\cite{zhan2021multi,an2020edge,zhang2020empowering, guan20185} are revolutionizing the way we think about information. These applications may consist of distributed devices that communicate with each other to perform a task. In particular, they may send information to each other. It is essential to keep the information fresh among them, because outdated information can degrade performance and even compromise human safety. 
	
	These applications can be made more flexible if their distributed devices communicate using a wireless network rather than a wired network\cite{huang2018new,liang2021design,hellstrom2019software}. Wireless networks obviate the need for installation and maintenance of wires. With a wireless network, the locations of the distributed devices can be changed in a moment without rewiring. Furthermore, many applications require the devices to be mobile and portable, and this is only possible with wireless networking. An application scenario is that of a central controller coordinating the movements of collaborating tetherless automatic guided vehicles (AGV)\cite{nakimuli2021deployment} via a wireless network.
	
	Many futuristic applications, such as industrial control applications, are time-sensitive in that they demand fast communication with low delays\cite{messenger2018time,luvisotto2019high,reill2018real,mathe2016control,toh2012performance}. Conventional wireless networks do not cater well to the time-sensitive requirements of these applications\cite{luvisotto2016ultra}. This work investigates a new wireless networking framework to meet the time-sensitivity requirements. We set the context of framework below.
	
	Generally, for networking purposes, a collection of networking services is used to support the communication between the distributed devices. In this paper, we refer to the collection of the networking services as the communication layer. The applications sitting at the application layer make use of the networking services of the communication layer to communicate with each other. The intricate interactions between the applications at the application layer and the networking services at the communication layer, and their ramifications for meeting the time-sensitivity requirements, have not been addressed in a deep and thorough manner in previous studies (see discussion in Section \ref{sec:relatedwork}). 
	
	\begin{figure}[!htbp]
		\centering
		\includegraphics[width=3in]{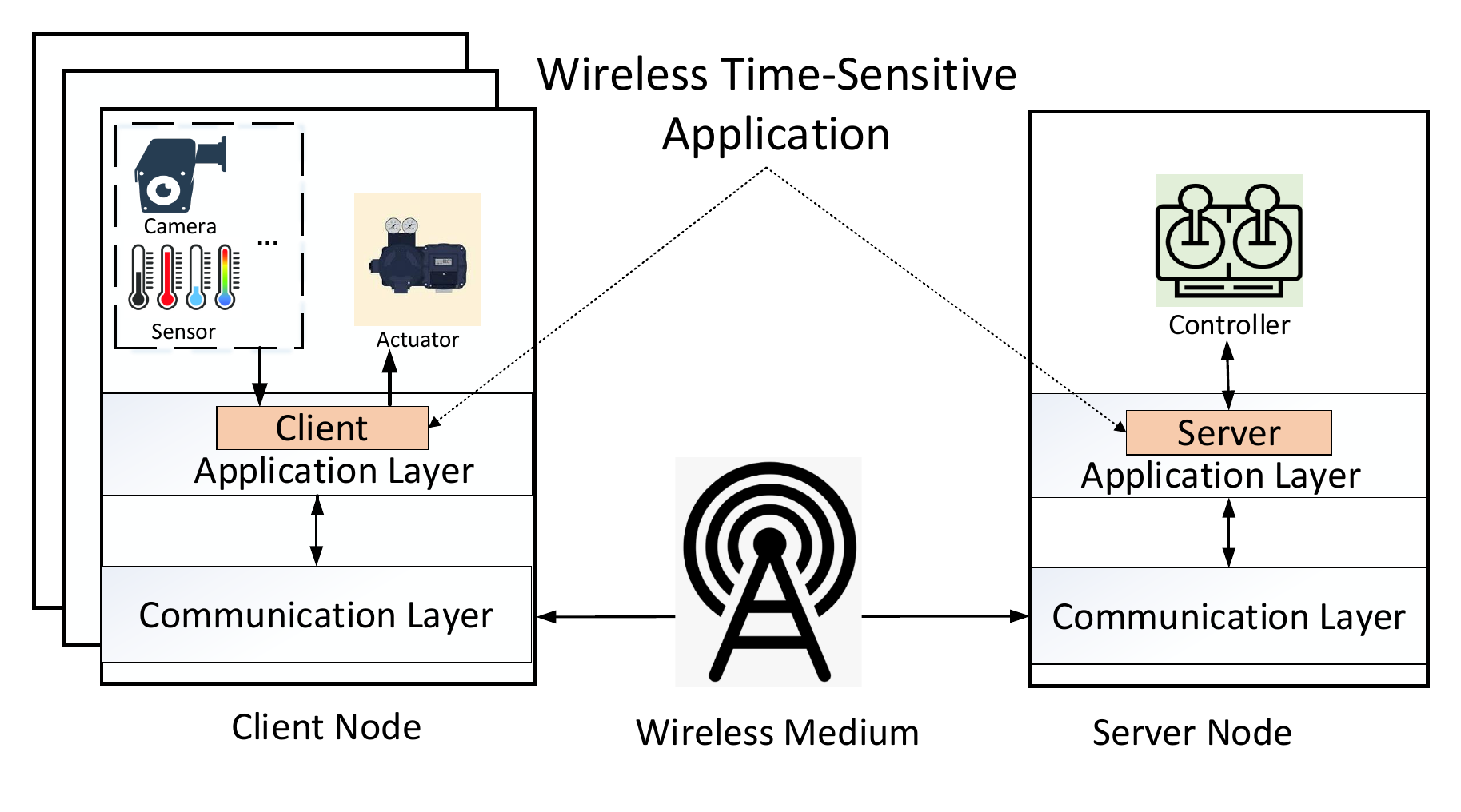} 
		\caption{An illustration of a wireless time-sensitive application.}
		\label{fig_1}
	\end{figure}
	
	This paper focuses on applications with a two-way closed-loop communication pattern. For applications with this communication pattern, a client communicates with a server in a request-response manner. Fig. \ref{fig_1} depicts a typical use case that involves the interactions of multiple client-server pairs over a shared wireless medium. In Fig. \ref{fig_1}, in the context of a client-server application, the sensor and actuator are the clients, and the controller is the server. To simplify exposition, let us assume that the sensor and the actuator are co-located at the same client, although in reality it does not have to be so. There is one single two-way communication session for each client-server pair. The two-way interaction is as follows: 
	\begin{enumerate}[1)]
		\item The client acquires some information (e.g., a measurement acquired from a sensor or an image captured from a camera) to generate a request message. It then forwards the request to its communication layer for delivery to the server. {The delivery time should be as short as possible once the measurement is made so that it remains fresh and updated when it reaches the server.}
		\item Upon receiving the request, the server, acting as a controller, formulates a response message based on the request. It then forwards the response message to its communication layer for delivery to the client, so that an actuator at the client can perform a certain operation based on the command embedded in the response. {The delivery time should be as short as possible once the response is generated so that the response is as fresh as possible when it reaches the client.}
	\end{enumerate}
	
	{This paper considers the “generate-at-will” model proposed in \cite{yates2015lazy}, where the sensor can sample the information about the observed phenomena at any time, and the request packet (or the so-called update packet) hence can be generated at any time of the user’s own choice. Such information update systems are expected to receive fresh updates from IoT devices in various Internet-of-Things (IoT) applications \cite{sun2019age,yates2021age}. For example, the “generate-at-will” model fits well with distributed control applications, where a sensor sends to a controller certain measurements in a periodic manner, and the controller, upon receiving each measurement, returns a control message. The overall feedback loop delay should be as low as possible for such control applications.}
	
	{To ensure the controller receives the freshest measurements, the sensor may orchestrate the periodic instants at which it makes the measurements so that the measurement instants match well with the instants at which transmission opportunities are available to the sensor. Similarly, at the controller side, if a transmission opportunity is available to the controller very soon after it has generated the control message, then the control message will remain fresh when it reaches the client.}
	
	{With the above in mind, we ask two fundamental questions that delve into the inner core of the time-sensitivity issue at hand:} 
	\begin{enumerate}[1)]
		\item When should the client acquire the information and generate the request so that it can be delivered by the communication layer to the server with minimal delays, taking into account the future instant when the communication layer may have a \textit{transmission opportunity}?
		\item When should the communication layer offer a \textit{transmission opportunity} to the server, subject to the networking construct, so that the server's response can be delivered to the client with minimal delays?
	\end{enumerate}
	
	We are interested in minimizing the request-response latency at the application layer – the time from the initiation of the request generation to the reception of the response at the client. We hereafter refer to this request-response latency as the \textit{application-to-application round trip time (RTT)}. We show via analysis and experiments that, without a mechanism to realize minimal round-trip delays between the client and server, no matter how fast and reliable the communication layer is, the application may still suffer from large and non-deterministic delays. In particular, tight coordination between the communication layer and the application layer is required to minimize the round-trip delay of the client-server interaction. 
	
	This paper puts forth two mechanisms to minimize the application-to-application RTT. Our mechanisms apply the following principles to address questions 1) and 2) above:
	\begin{enumerate}[]
		\item \textbf{\textit{{JIT Principle A}}}: The client application should initiate the generation of the request just before its communication layer has a transmission opportunity, not too early and not too late. 
		\item \textbf{\textit{{JIT Principle B}}}: The communication layer at the server should offer a transmission opportunity to the application layer just after the server has formulated and generated the response, not too early and not too late.
	\end{enumerate}
	
	We refer to the two principles as the Just-in-Time (JIT) principles and to a networking system that realizes the two principles as a JIT system. For concreteness, this paper assumes a time-division-multiple-access (TDMA) wireless network, in which time is divided into time slots and a message is transmitted in a time slot\cite{rubin1979message}. Thus, the aforementioned clients and servers make use of the time slots to communicate with each other at the communication layer, and \textit{time slots} correspond to the aforementioned \textit{transmission opportunities}.
	
	The main reason for assuming TDMA is that the physical-network latency can be made to be deterministic by preallocating time slots to users. For this reason, TDMA networks are often assumed to be used in time-critical networking\cite{cheng2017det,fernandez2019analysis,chang2016slot}. Although this work assumes the underlying network to be a TDMA network, it does not mean JIT principles are not viable if TDMA networks are not adopted. In Appendix C, we discuss what if a Carrier Sense Multiple Access (CSMA) network (e.g., Wi-Fi) is used instead. Essentially, we will need to replace the deterministic JIT here with a probabilistic JIT given that CSMA networks do not preallocate transmission opportunities to the nodes.

	To realize \textit{\textbf{JIT principle A}}, we design a \textit{\textbf{JIT-triggered packet generation}} mechanism in which the client is informed by a JIT middleware to generate a request slightly in advance of the transmission opportunity. It takes a certain amount of time for the client to execute a function to generate the request. Informing the client too late will cause the request to miss its transmission opportunity and informing the client too early will cause the request to age (i.e., the information in it becomes stale) while the request sits in the communication layer waiting for its transmission opportunity. To do its job, the JIT middleware, designed to sit in between the application layer and the communication layer, needs to synchronize the timings at the application layer and the communication layer and informs the client at just the right time. 
	
	To realize \textit{\textbf{JIT principle B}}, we design a \textit{\textbf{JIT time-slot allocation}} mechanism that allocates a time-slot pair to each client-server pair: one time slot for the client request and one time slot for the server response. It takes a certain amount of time for the server to generate the response upon receiving a request. Thus, the two time slots must be separated by just the right interval. If the server time slot occurs too early after the client time slot, the response at the server may arrive at its communication layer after the server time slot has transpired. On the other hand, if the server time slot is way past the client time slot, the response may arrive at the communication layer well before its transmission opportunity, inducing extra waiting time at the communication layer. In particular, the separation between the two time slots must be commensurate with the server's processing time.
	
	Our JIT system is implemented over the Openwifi project\cite{jiao2020openwifi} on a System-on-Chip (SoC) evaluation kit. This \textit{real-time} implementation enables the overall end-to-end application-to-application communication.  All the wireless signal processing and the communication between protocol stacks are executed in real-time, not offline. Our JIT implementation can serve as a useful experimental testbed for a whole host of wireless time-sensitive applications. Also, although our JIT implementation is over a special-purpose SoC evaluation kit, the JIT principles as expounded in this paper can be implemented over other platforms, including general-purpose computers.
	
	Experimental results indicate that our JIT system: i) can ensure information freshness of messages; ii) can achieve minimum application-to-application RTT; iii) is robust against factors — preemption delay in multitasking software and clock asynchronization between the application layer and the communication layer — that may compromise timing performance.
	
	The rest of this paper is organized as follows. Section \ref{sec:relatedwork} discusses related works. Section \ref{sec:motivation} presents the motivation for our work. Section \ref{sec:systemdesign} details the JIT system design. Section \ref{sec:implementation} and \ref{sec:Experiment} present our implementation and experimental results, and Section \ref{sec:conclusion} concludes this work.
	
	\section{Related Work} \label{sec:relatedwork}
	\subsection{Work Related to JIT principle A}
	Two related works are \cite{bartols2011performance}, \cite{liang2021design}. The communication layer of \cite{bartols2011performance} attaches a timestamp to the head-of-line packet of a queue within the communication layer just before sending it out. The “just-in-time” timestamp facilitates the measurement of the end-to-end latency incurred by the communication layer. Unlike our work here, \cite{bartols2011performance} does not have a mechanism to prevent or minimize aging of packets. In particular, the approach in \cite{bartols2011performance} does not generate the packet in a just-in-time manner. The packet is generated without regard to the transmission opportunity at the communication layer and it sits in a queue in the communication layer waiting for its transmission opportunity. It is the time stamp added to the packet that is “just-in-time”, in order that end-to-end communication delay can be measured accurately. This communication delay is not the actual delay as perceived by the application, because it does not include the delay incurred by the packet while sitting in the queue.

	{The work in \cite{liang2021design} shares the same spirit as our JIT mechanism in that a “packet” is prepared just before its scheduled transmission.} However, what are really prepared just before the transmission in \cite{liang2021design} are the baseband samples for the software-defined radio (SDR) platform rather than data generated at the application layer. In other words, the data may have been generated at the application layer a while ago, waiting to be converted to baseband samples for radio transmission. During this waiting time, the data may continue to age. By contrast, the mechanism propounded by our current paper enables the preparation of the application-layer data (e.g., taking a sensor reading) just before a transmission opportunity arises at the lower layer, significantly reducing the aforementioned waiting time. The challenge is that the just-in-time mechanism necessitates tight and timely coupling between layers, and to facilitate that, we implement our mechanism on the SoC platform.

	\subsection{Work Related to JIT principle B}
	The time-slot allocation problem in this paper is a TDMA scheduling problem of a single-hop network, where all nodes are within each other's communication range \cite{djukic2008delay,shen2013prioritymac,nakashima2017cross}. 
	
	Unlike in \cite{djukic2008delay,shen2013prioritymac,nakashima2017cross}, however, we consider a problem in which nodes are grouped into client-server pairs. Two directional links associated with a client-server pair need to be allocated a pair of time slots such that the time spacing between the time slots are large enough to accommodate the server processing delay and yet small enough as not to incur excessive delay. Specifically, the time spacing from the client transmission time slot to the server transmission time slot has to be no less than the processing delay required by the server to send a response back to the client upon a client's request. Any time spacing beyond the server processing delay will cause additional unnecessary round-trip delay. The time slots in between the two time slots of a client-server pair can be allocated to other client-server pairs. 
	
	Existing works \cite{djukic2008delay,shen2013prioritymac,nakashima2017cross}, on the other hand, did not take the server’s processing delay into account and hence do not have the minimum-spacing constraint in their problem formulations. Therefore, their scheduling methods can only ensure that the time slots of the two links are non-overlapping without regard to the spacing between them. 
	
	In the context of graph theory, the time-slot allocation problem in our JIT principle B can be regarded as a vertex arrangement problem, where links are modeled as vertexes. A client-server pair (the associated links) are modeled as adjacent vertexes. And the goal is to find an arrangement such that each pair of adjacent vertices satisfies a pre-defined minimum separation requirement (see Section \ref{sec:systemdesign}-2 for details). Two related NP-complete problems have been previously studied: the cycle-separation problem \cite{leung1984some} and the directed separation problem \cite{han1992scheduling}. The critical difference between our problem and the above two problems lies in the definition of the distance of adjacent vertices. This difference fundamentally changes the nature of our problem, and the solution to our problem cannot be found by the approaches of \cite{leung1984some} and \cite{han1992scheduling}. We will further elaborate the fundamental differences between our problem and those of \cite{leung1984some} and \cite{han1992scheduling} in Section \ref{sec:systemdesign}-2.
	\section{Motivation for JIT: A Quantitative Overview} \label{sec:motivation}
\begin{figure}[!htbp]
		\centering
		\includegraphics[width=3in]{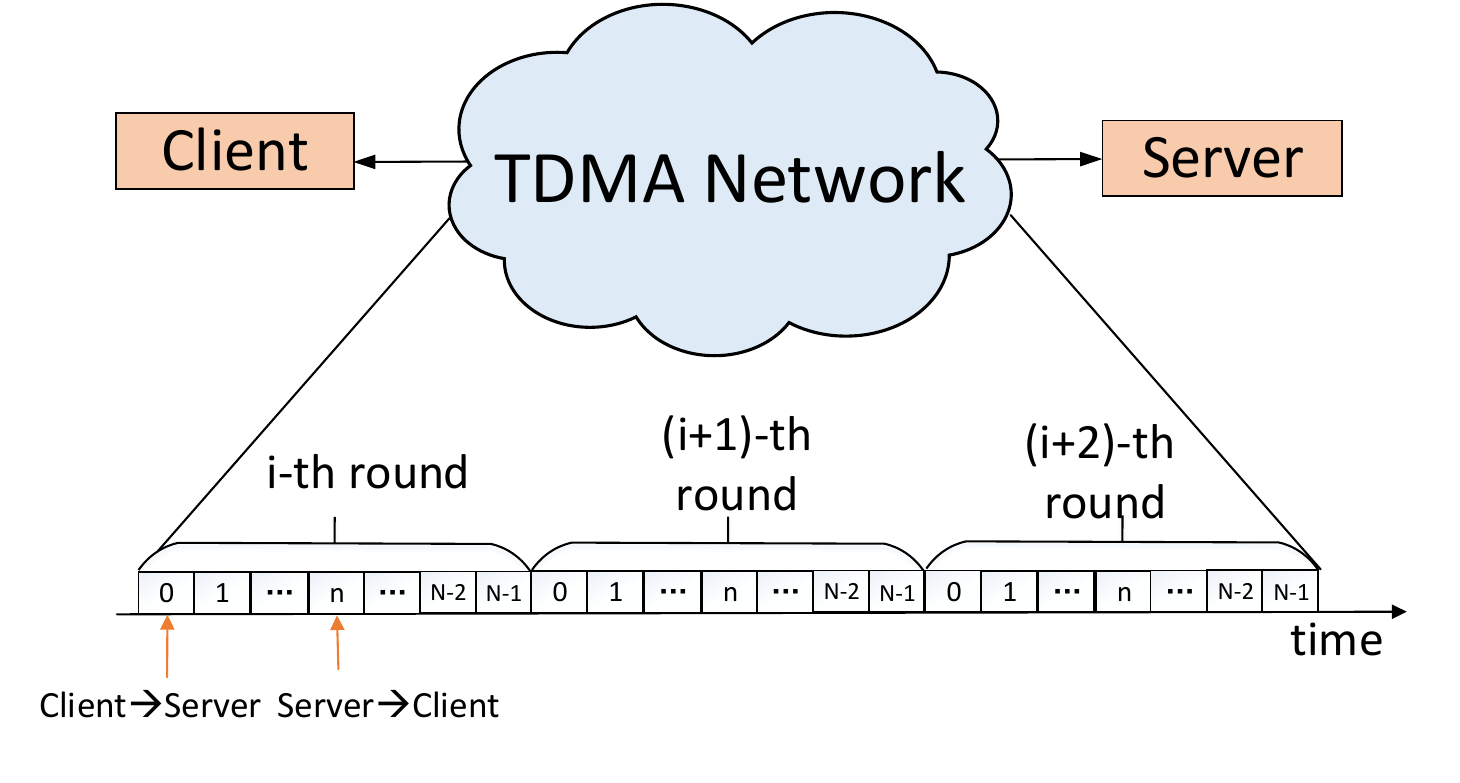}
		\caption{A client-server application over a TDMA network.}
		\label{fig_2}
	\end{figure}

	With respect to the use case depicted in Fig. \ref{fig_1}, let us trace through the delay components that contribute to the overall the application-to-application RTT when the application is deployed over a TDMA network. As illustrated in Fig. \ref{fig_2}, suppose that each TDMA round consists of $N$ time slots. Without loss of generality, in this example, time slot $0$ of all rounds is dedicated to the transmission of a request from the client to the server, and a time slot $n, n \neq 0$ of all rounds is dedicated to the transmission of a response from the server to the client. 
	
\begin{figure}[!htbp]
	\centering
	\includegraphics[width=3.7in]{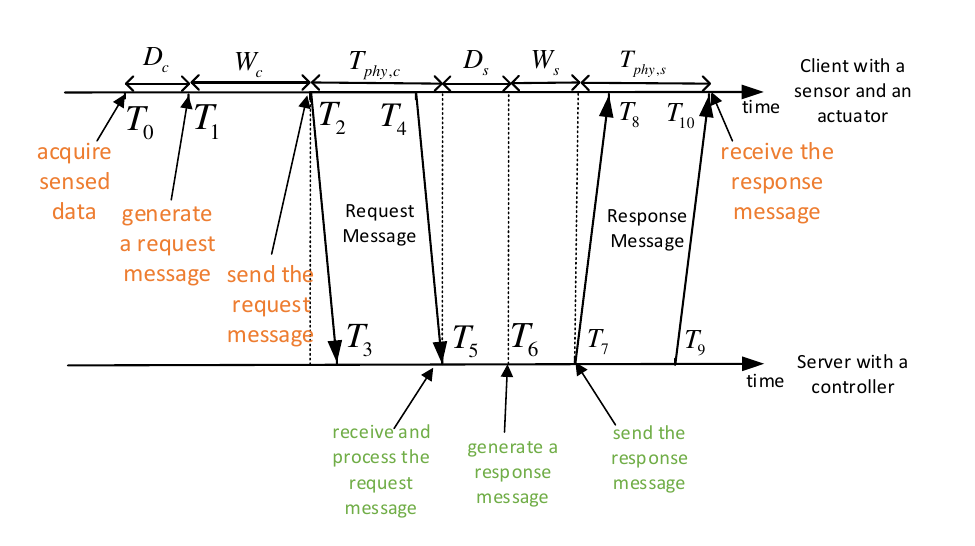}
	\caption{Time incurred by a sensor–controller–actuator feedback loop.}
	\label{fig_3}
\end{figure}

	To illustrate how a conventional networking system may not do the best job in minimizing the application-to-application RTT, let us consider the sequence of events in the feedback loop. Fig. \ref{fig_2} shows a typical time diagram of a sensor-controller-actuator feedback loop over the TDMA network:
	\begin{enumerate}[1)]
		\item With reference to Fig. \ref{fig_3}, the client acquires the sensed data from the sensor at time ${T_0}$ and generates a request ${M_c}$ at time ${T_1}$. Thus, ${D_c} = {T_1} - {T_0}$ is the time required to generate ${M_c}$. We refer to ${D_c}$ as the \textit{client's processing delay}.
		\item The client waits until ${T_2}$, the beginning of slot $0$ of the next upcoming TDMA round to transmit ${M_c}$ to the server, inducing an extra \textit{client's waiting time} of ${W_c} = {T_2} - {T_1}$. The transmission of ${M_c}$ takes a certain amount of transmission time, inducing a \textit{client's transmission delay}. In Fig. \ref{fig_3}, the transmission time or delay is ${T_4} - {T_2}$, where ${T_2}$ is the time at which the client transmits the first bit of ${M_c}$ and ${T_4}$ is the time at which of the client transmits the last bit of ${M_c}$.
		\item In addition, there is a \textit{client's propagation delay} from the client to the server. The server receives the first bit of ${M_c}$ at ${T_3}$, and receives the last bit of ${M_c}$ at ${T_5}$. We have ${T_5} - {T_3} = {T_4} - {T_2}$, and the propagation delay is ${T_3} - {T_2} = {T_5} - {T_4}$. We note that the physical medium is occupied for a duration of $T_{phy,c} = {T_5} - {T_2}$ after taking into account the propagation delay.
		\item Upon receiving ${M_c}$, the server takes some time to process it and to produce the control command as the response ${M_s}$, thus incurring a \textit{server's processing delay}. Suppose that the controller generates ${M_s}$ at ${T_6}$. Then the server's processing delay is ${D_s} = {T_6} - {T_5}$. 
		\item The server waits until ${T_7}$, the beginning of the time slot $n$ of a TDMA round, to transmit ${M_s}$ to the the client, inducing an extra \textit{server's waiting time} of ${W_s} = {T_7} - {T_6}$. 
		\item The first bit of response ${M_s}$ arrives at the actuator at ${T_8}$, and the last bit of ${M_s}$ arrives at the actuator at ${T_{10}}$. Thus, ${T_9} - {T_7} = {T_{10}} - {T_8}$ is the \textit{server's transmission delay} and ${T_8} - {T_7} = {T_{10}} - {T_9}$ is the \textit{server's propagation delay}. The physical medium occupied time is then ${T_{phy,s}} = {T_{10}} - {T_7}$. 
	\end{enumerate}

	Thus, the overall application-to-application RTT is made up of a number of delay components as written below:
	\begin{equation}
		\label{eqn:1}
		RTT = {T_{10}} - {T_0} = {D_c} + {W_c} + {T_{phy,c}} + {D_s} + {W_s} + {T_{phy,s}}.
	\end{equation}
	To reduce the application-to-application RTT, we need to reduce the extra waiting times ${W_c}$ and ${W_s}$. Let 
	\begin{equation}
		\label{eqn:2}
		W = {W_c} + {W_s}
	\end{equation}
	be the total extra waiting times. In the following, we argue that ${W_c}$ and ${W_s}$ can be large in a conventional networking system.
	
	Let ${t_{k,i}}$ be the beginning of time slot $k$ of round $i$. Recall that time slot $0$ of all rounds is dedicated to the client's transmission, and time slot $n$ of all rounds is dedicated to the server's transmission. Let us first consider ${W_c}$. Suppose that at time ${T_0} < t_{0,i}$, the client acquires a sensed data. The client then generates the corresponding request message ${M_c}$ that contains the sensed data at ${T_1}$. Two cases are possible: 
	\begin{enumerate}
		\item Suppose that ${\rm{ }}{T_1} \le t_{0,i}$. Then the client uses its assigned time slot in the round $i$ to transmit ${M_c}$. Then ${T_2} = t_{0,i}$, and we have 
		\begin{equation}
			\label{eqn:3}
			{W_c} = t_{0,i} - {T_1}.
		\end{equation}
		\item Suppose that ${T_1} > t_{0,i}$. Then the client misses its assigned time slot in the round $i$. The client needs to wait for time slot $0$ in TDMA round $i+1$ to transmit ${M_c}$. Thus, ${T_2} = t_{0,i + 1}$, and we have
		\begin{equation}
			\label{eqn:4}
			{W_c} = t_{0,i + 1} - {T_1} = F + t_{0,i} - {T_1},
		\end{equation}
		where $F$ is the duration of one round. From \eqref{eqn:4}, we can see that a larger $F$ will induce a larger ${W_c}$, leading to a larger application-to-application RTT. The worst-case maximum possible ${W_c}$ is obtained when ${T_1} = t_{0,i} + \varepsilon $ where $\varepsilon > 0$ is a tiny quantity. In this case, the waiting time is effectively $F$:
		\begin{equation}
			\label{eqn:5}
			{W_c} = F + t_{0,i} - t_{0,i} - \varepsilon \approx F.
		\end{equation}
	\end{enumerate}
	
	Let us next consider the maximum possible value of the total extra waiting time $W$ in \eqref{eqn:2}, denoted by ${W^*}$. In this case, we have case 2) in the above. Thus, the server receives ${M_c}$ at ${T_5}$ in round $i + 1$. It takes ${D_s}$ amount of time to process ${M_c}$ to generate the response message ${M_s}$ at ${T_6}$. The server then waits until the beginning of the next time slot $n$, ${T_7}$, to transmit ${M_s}$. As in the analysis of ${W_c}$, there are two possible cases:
	\begin{enumerate}[1)]
		\item Suppose that ${T_6} \le {t_{n,i + 1}}$. Then the server uses time slot $n$ of round $i + 1$ to transmit ${M_s}$. Thus, ${T_7} = {t_{n,i + 1}}$, and we have 
		\begin{equation}
			\label{eqn:6}
			{W_s} = {t_{n,i + 1}} - {T_6}.
		\end{equation}
		\item Suppose that ${T_6} > {t_{n,i + 1}}$. Then the server misses time slot $n$ of round  $i + 1$. Hence, the server needs to wait until time slot $n$ of round $i + 2$ to transmit ${M_s}$ (here we assume that ${T_5}{\rm{ + }}{D_s} = {T_6}  \le {t_{n,i + 2}}$). Then ${T_7} = {t_{n,i + 2}}$, and we have 
		\begin{equation}
			\label{eqn:7}
			{W_s} = {t_{n,i + 2}} - {T_6} = F + {t_{n,i + 1}} - {T_6}.
		\end{equation}	
		The worst-case maximum possible ${W_s}$ is obtained when ${T_6} = {t_{n,i + 1}} + \varepsilon$, where $\varepsilon > 0$ is a tiny quantity. In this case, we have 
		\begin{equation}
			\label{eqn:8}
			{W_s} = F + {t_{n,i + 1}} - {t_{n,i + 1}} - \varepsilon \approx  F.
		\end{equation}	
	\end{enumerate}
	
	From \eqref{eqn:3}, \eqref{eqn:4}, \eqref{eqn:6} and \eqref{eqn:7}, we can see that in a conventional network system, ${W_c}$ and ${W_s}$ are varying quantities, and depending on the situation, the application-to-application RTT could be large. From \eqref{eqn:5} and \eqref{eqn:8}, when ${T_1} = t_{0,i} + \varepsilon$ and ${T_6} = {t_{n,i + 1}} + \varepsilon$, the system incurs the maximum total extra waiting time ${W^*}$ given by
	\begin{equation}
		\label{eqn:9}
		{W^*} \approx 2F.
	\end{equation}
	
	From \eqref{eqn:9}, we can see that the conventional client-server applications can incur an extra delay of around $2F$ in the overall application-to-application RTT in the worst case. Motivated by the observation encapsulated in \eqref{eqn:1} and \eqref{eqn:9}, we propose the JIT system to reduce the total extra waiting time delay ${W^*}$. Our JIT system has two main features:
	\begin{enumerate}[1)]
		\item \textbf{JIT-triggered packet generation:} With this mechanism, the client can acquire the sensed data and generate ${M_c}$ just slightly before the communication layer offers a transmission opportunity at ${T_2}$. For example, consider the transmission opportunity at round $i$. Our JIT system notifies the client to start sensing at time ${T_0} = {t_{0,i}} - {D_c} - \varepsilon $ such that ${M_c}$ can be generated at time ${T_1}$ to satisfy  
		\begin{equation}
			\label{eqn:10}
			{T_1} = {t_{0,i}} - \varepsilon,
		\end{equation}
		where $\varepsilon $ is a small positive slack. This gives the minimum possible value of 
		\begin{equation}
			\label{eqn:11}
			{W_c} = \varepsilon.
		\end{equation}
		\item \textbf{JIT time-slot allocation:} Once the server receives ${M_c}$ and then generates ${M_s}$, it can send out ${M_s}$ immediately. This feature requires our JIT system to know the server's processing delay ${D_s}$ in advance of the time-slot allocation. For example, suppose that in the round $i$, the controller receives ${M_c}$ at  ${T_5}$, generates ${M_s}$ at ${T_6}$, and waits until ${T_7}$ to transmit ${M_s}$. Our system can assign a time slot to the server so that we have  
		\begin{equation}
			\label{eqn:12}
			{T_6} = {T_7} - \sigma,
		\end{equation}
		where $\sigma $ is the duration of one time slot in the worst case. This gives 
		\begin{equation}
			\label{eqn:13}
			{W_s} = \sigma.
		\end{equation}
	\end{enumerate}
	
	From \eqref{eqn:11} and \eqref{eqn:13}, we can then bound $W$ to 
	\begin{equation}
		\label{eqn:14}
		W' = \varepsilon  + \sigma,
	\end{equation}
	which can be much smaller than the ${W^*} \approx 2F$ in \eqref{eqn:9}. Importantly, the $W'$ in \eqref{eqn:14} is independent of the duration of a TDMA round, $F$, which could be large for a TDMA system supporting a large number of IoT devices with each device taking up one slot of a round. Recall the JIT principle A and the principle B in Section \ref{sec:introduction}. We emphasize that the JIT-triggered packet generation and the JIT time-slot allocation are the mechanisms that apply the JIT principle A and the JIT principle B, respectively. In this way, our JIT system with the above two features can be free of additional system delay, achieving minimum application-to-application RTT for client-server applications. 
	
	\section{System Design}	\label{sec:systemdesign}
\begin{figure}[!htbp]
		\centering
		\includegraphics[width=3.5in]{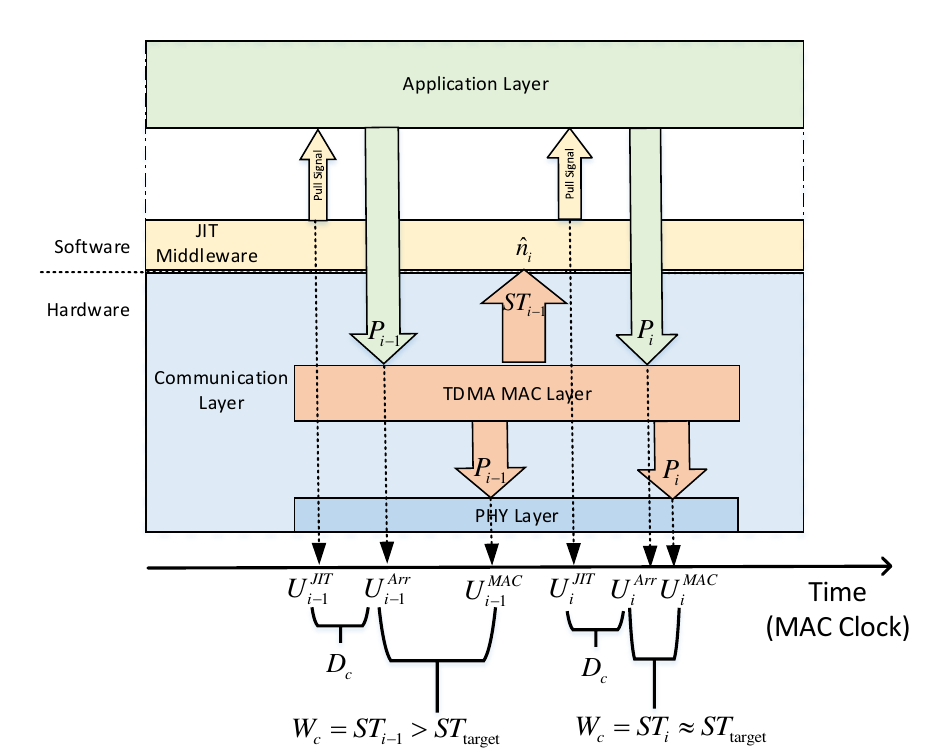}
		\caption{The architecture of the adaptive synchronization mechanism.}
		\label{fig_5}
	\end{figure}

	This section presents our JIT system design. The JIT system's network protocol stack is shown in Fig. \ref{fig_5}. The network protocol stack is partitioned into two main parts running on hardware and software, respectively. The communication layer, consisting of the MAC layer and the PHY layer, is implemented on hardware for high-speed processing and accurate timing control. The JIT middleware and the application layer are implemented in software. In particular, the JIT middleware, residing in between the application layer and the MAC layer, enables JIT-triggered packet generation at the client. Section \ref{sec:JIT-triggered packet generation} elaborates the JIT-triggered packet generation. And the JIT time-slot allocation for client-server applications is detailed in Section \ref{sec:JIT time-slot allocation in TDMA networks}.
	
	\subsection{JIT-triggered packet generation} 	\label{sec:JIT-triggered packet generation}
	This subsection presents the design of JIT-triggered packet generation at the client. A possible but na\"ive design is as follows: i) the MAC layer sends a pull signal to the JIT middleware; 2) upon receiving the pull signal, the JIT middleware relays the pull signal to the application layer, asking for a packet; 3) upon receiving the pull signal, the application layer starts to acquire sensed data and then generates the packet; 4) after being generated, the packet is sent to the JIT middleware immediately.
	
	This na\"ive design, however, has two limitations. First, the MAC layer initiates the pull signal and the JIT middleware relays the pull signal to the application layer. This relaying of the signal may induce extra delay. Second, since it takes time for the client's application to generate a packet, the MAC layer needs to know how much ahead of the assigned time slot it should send the pull signal to the JIT middleware to achieve accurate timing. Specifically, in this design, the MAC layer needs to have explicit knowledge of relaying delay incurred at the JIT middleware plus the client's processing delay.

	To circumvent the limitations, we devise a design for JIT-triggered packet generation as follows:
	\begin{enumerate}[(i)]
		\item Only the JIT middleware initiates a pull signal to the application layer to ask for a packet. The MAC layer does not send a pull signal to the JIT middleware. 
		\item A separate synchronization mechanism is used to synchronize the timing at the JIT middleware to the timing at the TDMA MAC layer. With the synchronization mechanism, the JIT middleware can initiate the pull signal at an appropriate time without explicit knowledge of the client's processing delay, such that the application-layer packet triggered by the pull signal can reach the TDMA MAC layer just in time for it to be transmitted. 
	\end{enumerate}
	
	We elaborate above mechanisms (i) and (ii) in the following. Before that, we emphasize that in a general set-up, the hardware clock used for communication may be different from the software clock used for computation. For example, in many computing devices that run applications, the applications are written in software and if they need to use timing functions (e.g., the code needs to sleep for a certain amount of time), they count on the operating system (OS) to provide these timing functions. The OS itself, in turn, makes use of the clock of the computing platform to offer the timing functions as software. Meanwhile, for communication and networking purposes, the communication platform (e.g., a Wi-Fi card attached to the computing device) uses its own internal oscillator\footnote{https://wiki.analog.com/resources/eval/user-guides/ad-fmcomms2-ebz/hardware/tuning\#fn\_\_1} (e.g., a voltage-controlled, temperature compensated oscillator) to provide the stable hardware clocking function for accurate communication timing. 
	
	Our JIT system is designed to cater to this general set-up. Specifically, we assume that the JIT middleware uses a software clock that is different from the hardware clock used by the TDMA MAC layer. We refer to the two clocks as “JIT clock” and “MAC clock”, respectively. We note that the JIT clock is the same clock as used by the client application software. For clarity, unless stated otherwise, all the time variables below are specified with reference to the MAC clock. Let $U_i^{MAC}$ denote the time at which the TDMA MAC layer is scheduled to begin to transmit the packet ${P_i}$. We have
	\begin{equation}
		\label{eqn:15}
		U_i^{MAC} = U_0^{MAC} + iF,i \ge 0,
	\end{equation}
	where $U_0^{MAC}$ is the time at which the TDMA MAC layer begins to transmit the first packet ${P_0}$ and $F$ is the duration of one TDMA round mentioned in Section \ref{sec:motivation}. Meanwhile, let $U_i^{JIT}$ denote the time when the JIT middleware is scheduled to pull packet ${P_i}$ from the application layer. We have
	\begin{equation}
		\label{eqn:16}
		U_i^{JIT} = U_0^{JIT} + iF',i \ge 0,
	\end{equation}
	where $U_0^{JIT}$ is the time when the JIT middleware sends the first pull signal to the application layer asking for the first packet ${P_0}$ and $F'$ is the duration of one TDMA round at the JIT middleware measured using the MAC clock. 
	
	Again, for exposition purposes, although the JIT middleware operates on the JIT clock, the time variables here, including $U_i^{JIT}$ and $F'$, are expressed in terms of the MAC clock, so that all variables use the same reference time. The reason that $F'$ and $F$ may not be the same is as follows. Suppose that the duration of one TDMA round is understood to be $F$ seconds. The MAC will implement the TDMA to have an exact duration of $F$ seconds according to its hardware clock. Meanwhile, the JIT middleware uses the software clock and also assumes the duration of one TDMA round is $F$ seconds. However, since the JIT layer uses the software clock for timing purposes, the resulting TDMA round at the JIT middleware may have a duration of $F'$ that is different from $F$ if $F'$ is expressed in terms of the MAC clock.  
	
	Note that we would like the differential $U_i^{MAC} - U_i^{JIT}$ to be not too small and not too large to allow sufficient time for the application to generate packet ${P_i}$ and then for ${P_i}$ to reach the TDMA MAC layer just in time for it to be transmitted by the MAC layer. 
	
	So far, all time variables have been specified with reference to the MAC clock. Now, since the operation at the JIT middleware only has access to the JIT clock, when we describe what the JIT middleware actually does in its operation, we need to assume the JIT middleware uses time variables with reference to the JIT clock. For clear notation, we put a “hat” over all time variables specified in the JIT clock. For example, for $U_i^{JIT}$ and $F'$ specified in MAC clock, $\hat U_i^{JIT}$ and $\hat F$ are their correspondences specified in the JIT clock. We further note that $\hat F = F$ because both the JIT middleware and the MAC layer have a common understanding of the duration of a TDMA round. Then, from the point of view of the JIT middleware, \eqref{eqn:16} can also be written as 
	\begin{equation}
		\label{eqn:17}
		\begin{array}{l}
			\hat U_i^{JIT} = \hat U_0^{JIT} + i\hat F {\kern 1pt} {\kern 1pt} {\kern 1pt} {\kern 1pt} {\kern 1pt} {\kern 1pt} {\kern 1pt} {\kern 1pt} {\kern 1pt} i \ge 1\\
			{\kern 25pt}{\rm{        =  }}{\kern 1pt}\hat U_{i - 1}^{JIT} + F.
		\end{array}
	\end{equation}
	
	To enable synchronized JIT-triggered packet generation, we need to address two issues arising from \eqref{eqn:15}, \eqref{eqn:16} and \eqref{eqn:17}:
	\begin{enumerate}
		\item \textbf{Clock Offset:} The JIT clock and the MAC clock have a relative clock offset that drifts with time because of the different tick rates of the two clocks. Now, define ${o_i}$ to be the offset between the above two clocks when the pull signal for ${P_i}$ is triggered. Specifically, we have
		\begin{equation}
			\label{eqn:18}
			\hat U_i^{JIT} = U_i^{JIT} + {o_i}.
		\end{equation}
		The difference $U_i^{MAC} - U_i^{JIT} = (U_0^{MAC} + iF) - (\hat U_0^{JIT} + iF - {o_i}) = (U_0^{MAC} - U_0^{JIT}) + ({o_i} - {o_0})$ is not a constant and it drifts with the term $({o_i} - {o_0})$. Let us further focus on $({o_i} - {o_0})$. 
		
		{From \eqref{eqn:18}, we have
		\begin{equation}
			\label{eqn:19}
			\begin{array}{l}
			{o_i} - {o_0}\\
			= (\hat U_i^{JIT} - \hat U_0^{JIT}) - (U_i^{JIT} - U_0^{JIT})\\
			= \int_{\hat U_0^{JIT}}^{\hat U_i^{JIT}} {d\hat t}  - \int_{U_0^{JIT}}^{U_i^{JIT}} {dt} 
			\end{array}
		\end{equation}
		where $\hat t$ and $t$ are the continuous time units of the JIT clock and the MAC clock, respectively.}
		
		{Now suppose that we expressed the relationship between $\hat t$ and $t$ by an increasing function $\hat t = f(t)$. Then continuing from the above, and by change of variables in the first integral,  we have 
		\begin{equation}
			\label{eqn:19-1}
			\begin{array}{l}
				{o_i} - {o_0}\\
				= \int_{{f^{ - 1}}(\hat U_0^{JIT})}^{{f^{ - 1}}(\hat U_i^{JIT})} {\frac{{df(t)}}{{dt}}} dt - \int_{U_0^{JIT}}^{U_i^{JIT}} {dt} \\
				= \int_{U_0^{JIT}}^{U_i^{JIT}} {\left( {\frac{{df(t)}}{{dt}} - 1} \right)} dt\\
				= \int_{U_0^{JIT}}^{U_i^{JIT}} {\left( {\frac{{d\hat t}}{{dt}} - 1} \right)} dt
			\end{array}
		\end{equation} 
		where ${\rm{ }}{f^{ - 1}}(\hat U_i^{JIT}) = U_i^{JIT}$ and ${f^{ - 1}}(\hat U_0^{JIT}) = U_0^{JIT}$, and we assume the derivative $\frac{{d\hat t}}{{dt}}$ is express in terms of $t$ as a function in the above. If the tick rate of the JIT clock is faster than the tick rate of the MAC clock, then $\frac{{d\hat t}}{{dt}} = \frac{{df(t)}}{{dt}} \ge 1$, and vice versa.} Thus, we can see from \eqref{eqn:19-1} that for an unsynchronized system where $\frac{{d\hat t}}{{dt}} \ne 1$, $U_i^{MAC} - U_i^{JIT}$ may grow unbounded or may become negative. For just-in-time operation, $U_i^{MAC} - U_i^{JIT}$ for all $i$ must be kept within a tight range. To achieve this, we need a synchronization mechanism. We will present our synchronization mechanism shortly.
		\item \textbf{Timing Imprecision:} Rather than having precise time control as in hardware, the multitasking software architecture in the operating system (OS) may cause the JIT middleware and the application layer to experience preemptions from other tasks. In general, a modern computer OS segregates virtual memory into user space and kernel space. Kernel space is strictly reserved for running a privileged operating system kernel, kernel extensions, and most device drivers. In contrast, user space is the memory area where application software and some drivers execute. Processes in user space have low priorities and can be preempted by processes in kernel space (e.g., system calls, process scheduling, memory management and inter-process communication).\footnote{ https://en.wikipedia.org/wiki/User\_space}
		As a result, ${D_c}$ could vary for different packets. Let $D_{c,i}$ denote the client's processing delay for packet ${P_i}$. In general, ${D_{c,i}}$ in an actual system includes not only the time for packet generation, but also the preemption delay caused by other tasks preempting our task plus the time for the packet to be delivered from the application layer to the MAC layer. A larger preemption delay induces a larger ${D_{c,i}}$. Note also that ${D_{c,i}}$ is unknown to the JIT system.
	\end{enumerate}
	Packet ${P_i}$ arrives at the MAC layer at time $U_i^{Arr} = U_i^{JIT} + D_{c,i}$. Let $S{T_i}$ denote the slack time for packet ${P_i}$ at the TDMA MAC layer. Specifically, 
	\begin{equation}
		\label{eqn:20}
		U_i^{MAC} = U_i^{Arr} + S{T_i},i \ge 0.
	\end{equation}
	In \eqref{eqn:20}, $ST_i$ is exactly the extra client’s waiting time $W_c$ that the JIT principle A aims to reduce, as discussed in Section \ref{sec:motivation}. Thus, we have two requirements:
	\begin{enumerate}[(i)]
		\item We want to minimize the slack time $S{T_i}$ as much as possible to allow just-in-time packet transmission. 
		\item We would like to have $S{T_i} \ge 0$, so that we can prevent ${P_i}$ from missing its desired transmission time at the MAC layer $U_i^{MAC}$.
	\end{enumerate} 
	
	\subsubsection{Synchronization Mechanism}
	To meet the requirement so that we have $S{T_i} \ge 0$, but not too much larger than $0$, we devise an adaptive synchronization mechanism to synchronize $U_i^{JIT}$ to $U_i^{MAC}$. This adaptive synchronization mechanism involves interaction between the JIT middleware and the MAC layer. 
	
	In place of \eqref{eqn:17}, the JIT middleware adjusts its transmission time for the $i$-th packet $\hat U_i^{JIT}$ as follows:
	\begin{equation}
		\label{eqn:21}
		\hat U_i^{JIT} = \hat U_{i - 1}^{JIT} + F + {\hat n_i},{\kern 1pt} {\kern 1pt} {\kern 1pt} {\kern 1pt} {\kern 1pt} {\kern 1pt} {\kern 1pt} {\kern 1pt} {\kern 1pt} i \ge 1,
	\end{equation}
	where ${\hat n_i}$ is a timing offset adjustment for the JIT pull signal for packet ${P_i}$ computed based on feedback from the MAC layer. Specifically, based on the transmission of ${P_{i - 1}}$, we determine ${\hat n_i}$ as follows:
	\begin{enumerate}
		\item \textbf{Obtain and feed back the slack time $S{T_{i - 1}}$ of the packet ${P_{i - 1}}$:} As illustrated in Fig. \ref{fig_5}, the JIT middleware sends a pull signal at time $\hat U_{i - 1}^{JIT}$ (w.r.t. the JIT clock) to the application layer. This triggers the generation of the packet ${P_i}$. Upon receiving packet ${P_i}$ from the application layer, the MAC layer retrieves the packet's scheduled transmission time $U_{i - 1}^{MAC}$ (w.r.t. the MAC clock) and obtains the slack time $S{T_{i - 1}}$(w.r.t. the MAC clock): 
		\begin{equation}
			\label{eqn:22}
			S{T_{i - 1}} = U_{i - 1}^{MAC} - U_{i - 1}^{Arr},i \ge 1,
		\end{equation}
		where $U_{i - 1}^{Arr} = U_{i - 1}^{JIT} + D_{c,i - 1}$ is the time when ${P_{i - 1}}$ arrives at the MAC layer. Note that although the MAC layer does not know $U_{i - 1}^{JIT}$ and ${D_{c,i - 1}}$, it does know $U_{i - 1}^{Arr}$ since this is the time the packet arrives at the MAC layer according to the MAC clock. The MAC layer then feeds back $S{T_{i - 1}}$ to the JIT middleware. 
		
		\item \textbf{Set the target slack time ${ST}_{\rm{target}}$ for all packets at initialization:} The JIT middleware aims to achieve a target slack time ${ST}_{\rm{target}}$ at the MAC layer for all packets. As a conservative measure to avoid underflow at the MAC layer, we would like to set ${ST}_{\rm{target}}$ to the maximum preemption delay that the packet-generation process could encounter, explained as follows. We aim for a set-up such that in the worst case, when the packet ${P_i}$ incurs the maximum preemption delay, it would still arrive at the MAC layer at time $U_i^{MAC}$, just in time for it to be transmitted. Thus, if $S{T_j}$ for different $j$ hovers around ${ST}_{\rm{target}}$, and if a particular packet ${P_i},i > j$, suddenly incurs a worst-case preemption delay, it would still arrive at the MAC layer in time for its transmission, even if all the prior packets ${P_j},j < i$, incur the minimum preemption delay.
		
		Since ${D_{c,i}}$ for different $i$ fluctuates with their instantaneous preemption delays, the maximum preemption delay is also the maximum time difference between the generations of two successive packets (other delay components being constant and the preemption delay being the only variable delay). To estimate the maximum preemption delay, the JIT middleware generates $Q$ packets during initialization before real packet transmissions. For each of the packet generated, we monitor its start time and the end time, and calculate its processing delay. Since we can only use the software clock to obtain the start time of a software process, the processing delay we calculate in this way is specified in terms of the JIT clock. Let ${\hat D_{c,q}}$ denote the calculated processing delay using the JIT clock. We derive $\hat{ST}_{\rm{target}}$ by 
		\begin{equation}
			\label{eqn:23}
			\hat{ST}_{\rm{target}} = \mathop {\max }\limits_{1 \le q \le {Q},1 \le r \le {Q},{\rm{ and }}~ r \ne q} ({\hat D_{c,q}} - {\hat D_{c,r}}).
		\end{equation}
		In particular, $\hat{ST}_{\rm{target}}$ approaches the maximum preemption delay when $Q$ is large. 
		
		\item \textbf{Determine the timing offset ${\hat n_i}$ for packet ${P_i}$:} With respect to the clock-offset issue, the JIT middleware needs to ensure that in the near future, $S{T_i}$ does not deviate from $\hat{ST}_{\rm{target}}$ too much. This is achieved by adjusting the timing offset ${\hat n_i}$ for packet ${P_i}$. Specifically, we adjust ${\hat n_i}$ using the following exponential smoothing formula acting as a low-pass filter to remove high-frequency jitters in the measurements so that the system does not overreact to short-term fluctuations:
		\begin{equation}
			\label{eqn:24}
			{\hat n_i} = (1 - \alpha ){\hat n_{i-1}} + \alpha (S{T_{i-1}} - \hat{ST}_{\rm{target}}), i \ge 1,
		\end{equation}
		where $0 < \alpha \le 1$ is a constant smoothing parameter. In \eqref{eqn:24}, the smoothed statistic $\hat n_i$ is a weighted average of the current observation $S{T_{i-1}} - \hat{ST}_{\rm{target}}$ and the previous smoothed statistic $\hat n_{i-1}$. {In general, if $\alpha$ is large, then the bandwidth of the low-pass filter is large (i.e., less smoothing), and if $\alpha$ is small, then the bandwidth of the low-pass filter is small (i.e., more smoothing). The main idea is that, by choosing an appropriate value for $\alpha$, we want to filter out short-term fluctuations while retaining the long-term drift so that our synchronization mechanism can react to the long-term drift.} If $\alpha$ is too large, our mechanism may overreact to short-term fluctuations due to “measurement noise”; on the other hand, if $\alpha$ is too small, our system may not be able to track the drift in a timely manner for tight synchronization. For our system, we set $\alpha = 0.6$. This value is found from experimentation to provide good tracking performance. For initialization, the JIT middleware sets $\hat n_0 = 0$ for the packet ${P_0}$.
	\end{enumerate} 
	After deriving ${\hat n_i}$, the JIT middleware calculates $\hat U_i^{JIT}$ using \eqref{eqn:21} and sends a pull signal at time $\hat U_i^{JIT}$ to the application layer, asking for the next packet ${P_i}$. 
	
	Note that in \eqref{eqn:24} $\hat{ST}_{\rm{target}}$ is specified in terms of the JIT clock. However, $S{T_i}$ in \eqref{eqn:23} is specified in terms of the MAC clock. Ideally, we would like them to be expressed in terms of the same clock, particularly when they are applied in \eqref{eqn:24}. We provide an analysis in Appendix \ref{appendix:I} to show that, despite having these two terms expressed in terms of different clocks, our adaptive synchronization mechanism is still stable. Furthermore,, the differential $U_i^{MAC} - U_i^{JIT}$ can be bounded as shown in Appendix \ref{appendix:II}.
	
	\subsection{JIT time-slot allocation in TDMA networks} 	\label{sec:JIT time-slot allocation in TDMA networks}
	We next consider issues related to time-slot allocation in a TDMA network when running the JIT system. Consider a server paired with ${N_c}$ clients in the TDMA network. As related in Section \ref{sec:motivation}, each client sends a request to the server. Upon receiving the request, the server sends back a response. Let ${C_j}$ be the  $j$-th client and $S$ be the server. For clarity, we use another notation ${S_j}$ to denote the server $S$ when it communicates with client ${C_j}$. Given a specific $j$, ${C_j}$ and ${S_j}$ together represent the $j$-th communication pair.
	
	Let $\Delta t$ be the duration of a time slot. We require that $\Delta t = \mathop {\max }_j \max (T_{phy,{c_j}},T_{phy,{s_j}})$, where $T_{phy,{c_j}}$ and $T_{phy,{s_j}}$ are the $j$-th communication pair's occupied airtime on the physical wireless medium from the client to the server and from the server to the client, respectively (see Section \ref{sec:motivation}). For simplicity, we assume the request and response messages are of the same size. We further assume that the differences in the propagation delays of different communication pairs are negligible. We could then write $T_{phy,{c_j}} = T_{phy,{s_j}} = {T_{PHY}}$, where ${T_{PHY}}$ is constant for different communication pairs. Thus, $\Delta t = {T_{PHY}}$. To simplify discussion in the following, let us assume that all times are specified in terms of the time unit of one time slot. Thus, $\Delta t = 1$. 
	
\begin{figure}[!]
		\subfigure[Clockwise time-slot progression.]{\label{fig_6:a}\includegraphics[width=1.3in]{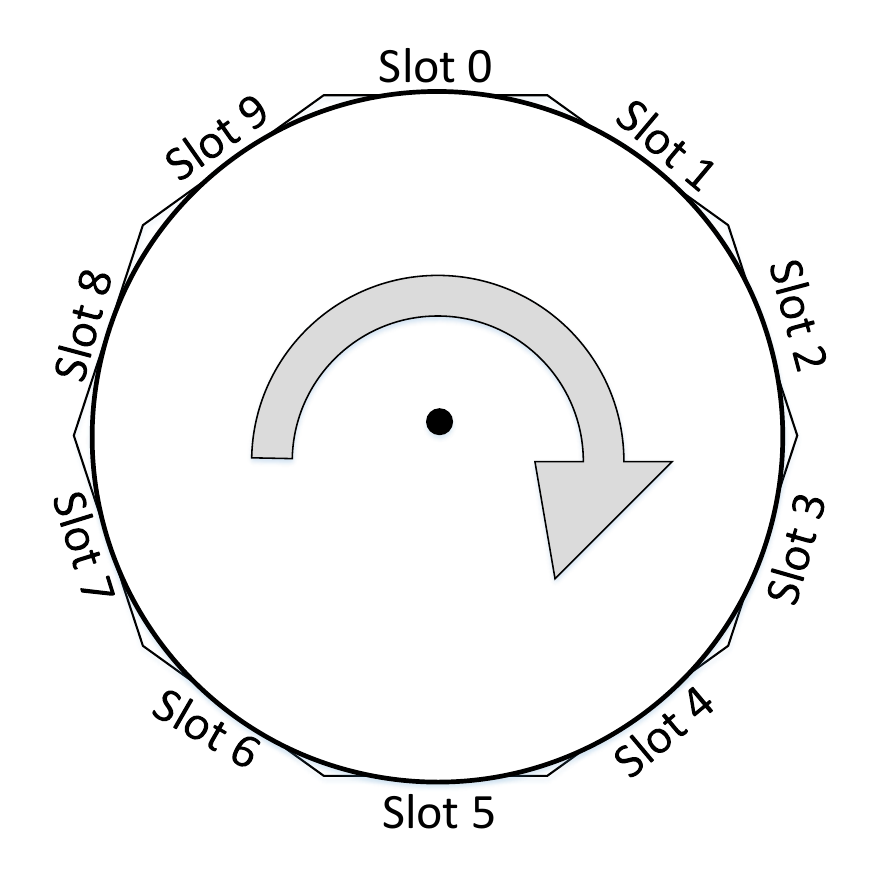}}
		\subfigure[Only one induced subring when $\beta = 3$.]{\label{fig_6:b}\includegraphics[width=2.0in]{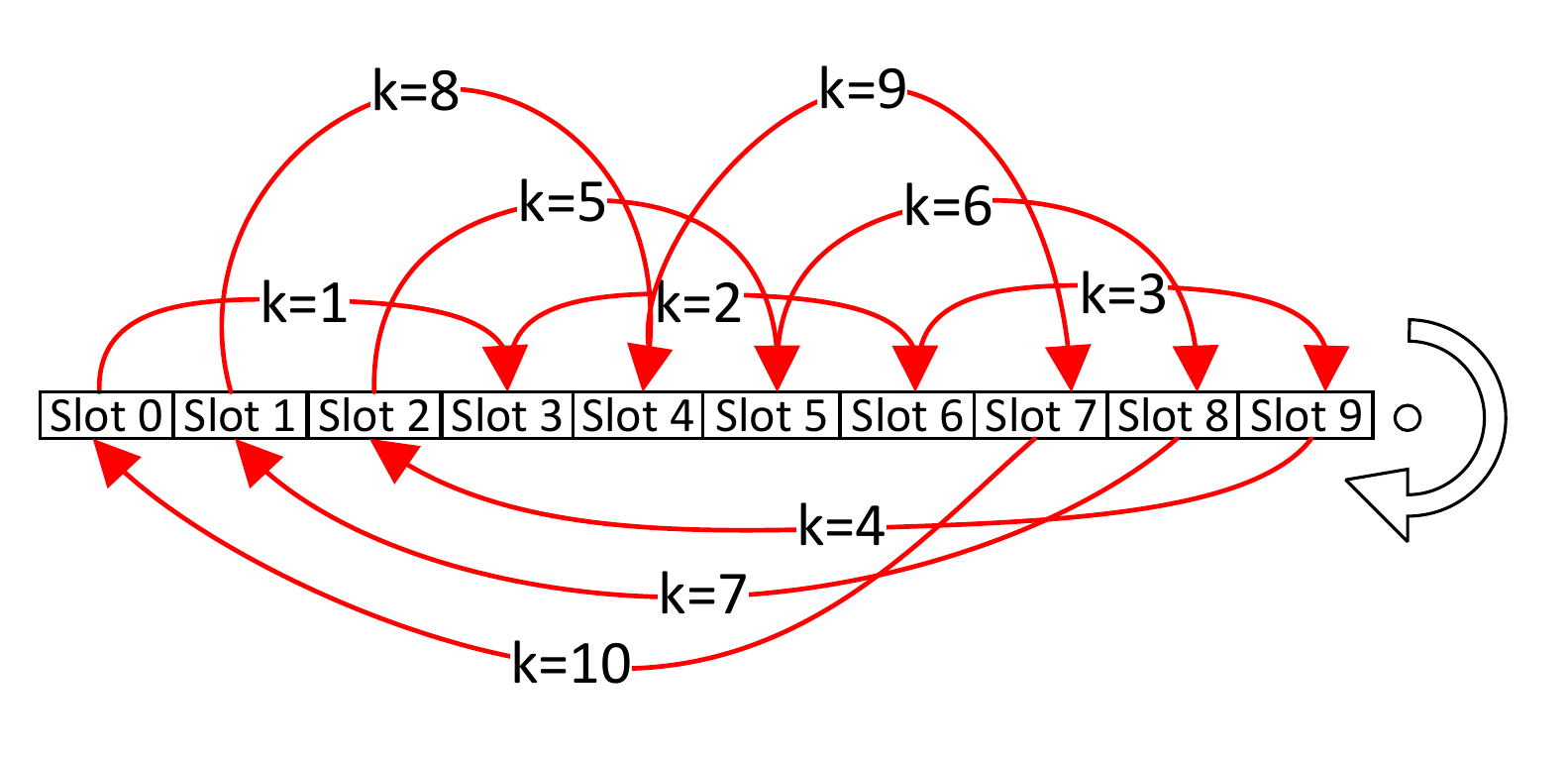}}
		\subfigure[Induced subrings will not yield optimal packing when $\beta = 2$.]{\label{fig_6:c}\includegraphics[width=2.4in]{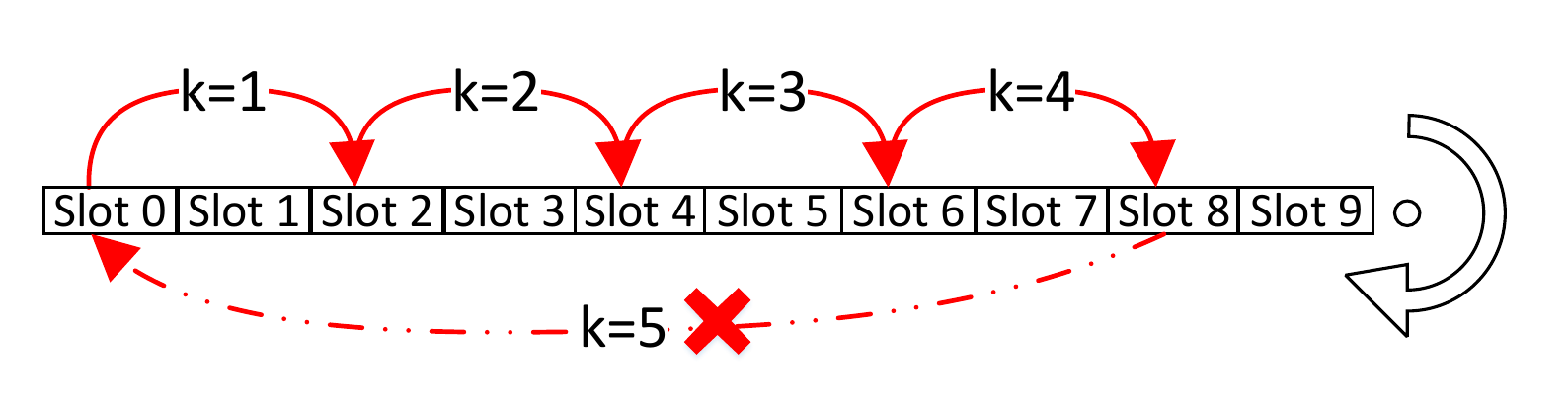}}
		\centering
		\subfigure[Five induced subrings when $\beta = 5$.]{\label{fig_6:d}\includegraphics[width=2.4in]{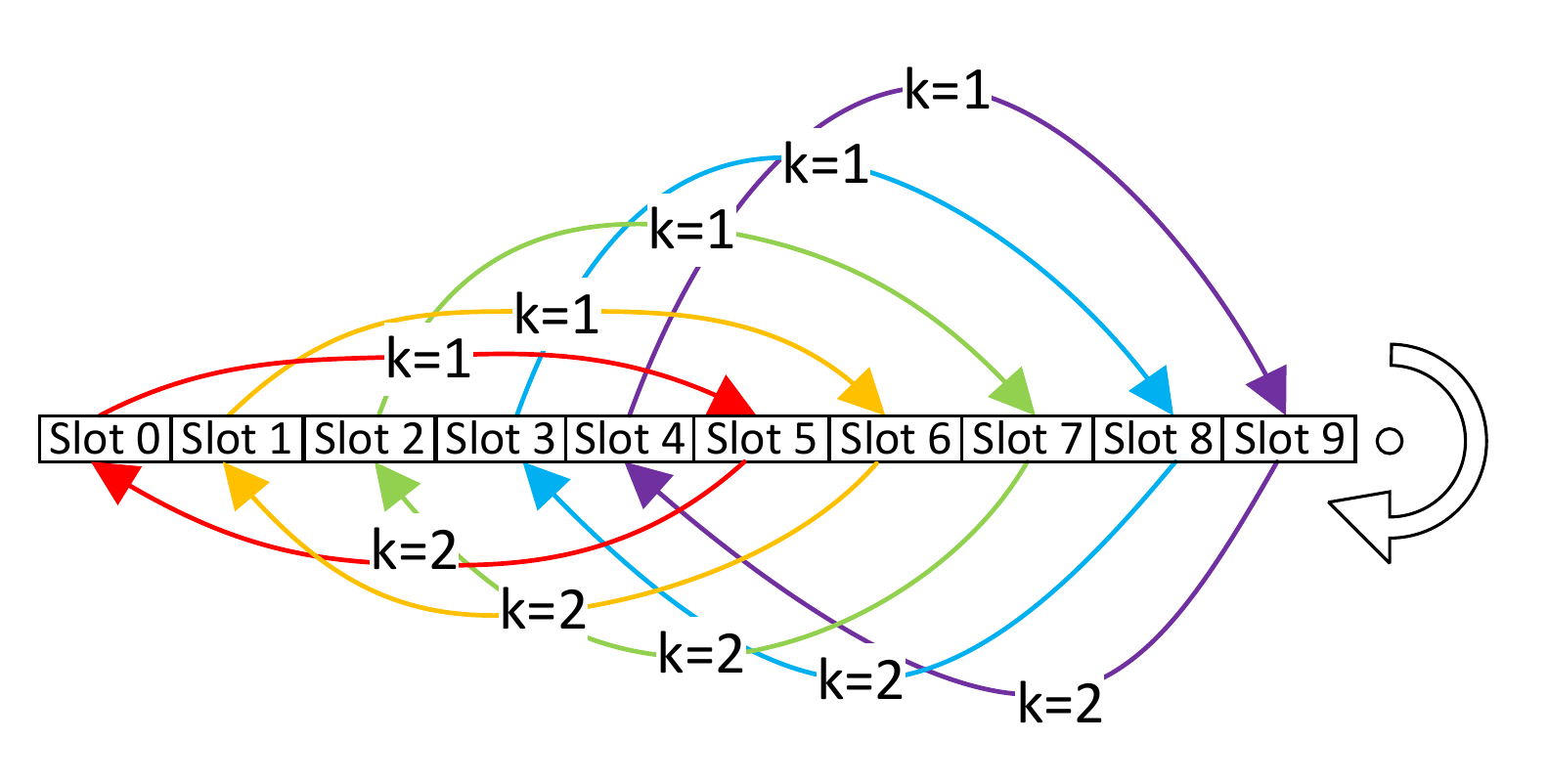}}
		\caption{Illustration of time-slot progression over a ring when $N=10$. In figures (b), (c), (d), we draw the ring as a line to avoide cluttering, with the implicit understanding that slot $9$ wraps back to slot $0$.}
		\label{fig_6}
	\end{figure}

	Let $TS({C_j}) \in \{ 0,...,N - 1\} $ and $TS({S_j}) \in \{ 0,...,N - 1\} $ be the time slots in TDMA rounds assigned to ${C_j}$ and ${S_j}$, where $N$ is the number of time slots per TDMA round. At issue is the following question: what should be the pair of time slots ${\{ (TS({S_j}),TS({C_j}))\} _{j = 0,1,...,{N_c} - 1}}$ allocated to the client-server pairs in order to minimize the response time of the client-server applications? To aid the understanding of the issue at hand, let us picture the progression of the time slots in successive rounds as a clockwise movement over a ring, as illustrated in Fig. \ref{fig_6} (a), where we assume $N = 10$. For example, slot $1$ follows slot $0$; slot $0$ follows slot $N-1$. Recall from Fig. \ref{fig_3} that the JIT principle B is to minimize $W_s = T_7 - T_5$ given the server processing delay $D_s$. With ${\{ (TS({S_j}),TS({C_j}))\}_j}$, the goal is equivalent to moving around the ring in a clockwise manner from $TS({C_j})$ to $TS({S_j})$ such that the number of time slots that has transpired is at least 
	\begin{equation}
		\label{eqn:25}
		{\beta _i} = \left\lceil {{D_{s,i}}} \right\rceil  + 1\bmod N.
	\end{equation}
	For the distance constraint  $\beta _i$ in \eqref{eqn:25}, we may have to cycle around the ring a number of times. For example, if $\left\lceil {{D_{s,i}}} \right\rceil  + 1 \ge N$, then the time slot $TS({S_i})$ must occur in a round later than the round in which time slot $TS({C_i})$ occurs.

	
	In the context of graph theory, our problem can be regarded as a vertex arrangement problem over indexes on a ring. Specifically, we can construct an undirected graph $G(V,E)$ whereby the vertexes are the client requests and the server responses. There is an undirected edge between the two vertexes corresponding to the client request-server response pair $j$, but there is no edge between vertices of different client request-server response pairs. The time slot assignment $TS:V \to \{ 0,1,...,N - 1\}$ is an injective function from the vertexes to the indexes (if there are $N/2$ client-server pairs – i.e., $N$ vertexes – then the function is a bijective function). The distance between a client request-server response pair under the injection is defined to be ${E_j} = [TS({S_j}) - TS(C_j)]\bmod N$. The objective of the vertex arrangement problem is to minimize the sum distance $\sum\nolimits_j {{E_j}} $. 
	
	{The authors in \cite{leung1984some}} studied a cycle-separation problem. As in our problem, the issue is arranging vertexes over a ring (cycle). However, the definition of distance is ${E_j} = \min (\left| {TS({S_j}) - TS(C_j)} \right|,N - \left| {TS({S_j}) - TS(C_j)} \right|)$. {In other words, there are two possible distances between the client and the server on the ring: the clockwise distance and the counter-clockwise distance between them. The investigation in \cite{leung1984some} defines the distance to be the smaller of the two.} For our problem, the distance is always the clockwise distance from the client to the server. This makes our problem fundamentally different from that of \cite{leung1984some}. 
	
		{The authors in \cite{han1992scheduling}} studied a directed separation problem where the arrangement of the vertexes is over a line rather than a ring. The distance is defined to be ${E_j} = TS({S_j}) - TS(C_j),TS({S_j}) \ge TS(C_j)$. Again, the different definition of the distance makes this problem fundamentally different from ours.

	In short, our problem is to construct a structural graph whereby the graph itself is already the multiple disjoint cycles we want without the need of deriving these cycles. In this way, our problem is not an NP-complete problem. Rather, it is a problem of identifying when such a graph will fulfill the distance-constraint requirement. For example, it is trivial to see that if each of the cycles has an even number of vertexes, then there is an optimal solution. By contrast, if each of the cycles has an odd number of vertexes, then an optimal solution is possible. As will be discussed in the following, our work identifies the condition leading to the possibility and impossibility of an optimal solution. We also provide a general method to construct the graph (and therefore the multiple cycles), and the complexity is of order $O(N)$.
	
	\subsubsection{Optimal Packing - A Special Combinatorial Problem}
	We next consider a combinatorial problem arising from our system. We assume that $N$ is even and that there are $N/2$ client-server pairs. For simplicity, we further assume that ${\beta _j} = \beta $ for all $j$. We ask the following question: Under what setting is it possible to have a bijection $TS:V \to \{ 0,1,...,N - 1\} $ such that ${E_j} = \beta, \forall j \in \{ 0,1,...,N - 1\} $? Note that if this is possible, we have an optimal solution in that the distance between each client-server pair $j$ meets the lower bound $\beta $ exactly. We refer to such an optimal solution, if possible, as \textit{\textbf{optimal packing}}. In the following, let us use the notation $(c,s)$ to denote the pair of time slots allocated to a client-server pair. For example, the following pairs give an optimal packing for the case of $\beta  = 3$ and $N = 10$: $\{ (0,3),(6,9),(2,5),(8,1),(4,7)\} $.
	
	\subsubsection{Optimal Packing Construction}
	We first discuss when optimal packing is possible. After that, we give a general method to construct optimal packing when it is possible. Consider time slots indexed by $0,1,...,N - 1$ on the ring, where $N$ is even. From any index $l,0 \le l \le N - 1$ on the ring, we can induce a subring by enumerating the indexes $l, (l + \beta )\bmod N, (l + 2\beta )\bmod N,...., (l + k\beta )\bmod N...$. Note that somewhere in this sequence, the index $l$ must repeat, since there is only a finite number of indexes on the ring. Let $k^*$ be the smallest integer such that $(l + k^*\beta) \bmod N = l$. The period of this subring is said to be $k^*$ and there are $k^*$ unique indexes on the sequence. 
	
	\paragraph*{Case 1} If the period $ k^* = N$, the subring is the full ring and optimal packing is possible. Fig. \ref{fig_6} (b) illustrates the full ring induced in the case of $l = 0$, $\beta  = 3$ and $N = 10$. Any $l \in \{ 0,...,9\} $ will induce the same full ring. With the full ring, we have found two possible optimal packings: $\{ [l, (l + \beta )\bmod N], [(l + 2\beta )\bmod N, (l + 3\beta )\bmod N], ..., [(l + (k^* - 2)\beta )\bmod N, (l + (k^* - 1)\beta )\bmod N{\kern 1pt} {\kern 1pt}]\} $ and $\{ [(l + (k^* - 1)\beta )\bmod N,l],[(l + \beta )\bmod N,(l + 2\beta )\bmod N],..., [(l + (k^* - 3)\beta )\bmod N,(l + (k^* - 2)\beta )\bmod N]\} $. The first packing is to have time slot $l$ as a client time slot, and the second packing is to have time slot $l$ as a server time slot. There is no other possible optimal packing. The reason is simple. If we make time slot $l$ a client time slot, then time slot $(l + \beta )\bmod N$ must be a server time slot. This means time slot $(l + 2\beta)\bmod N$ must be a client time slot, and so on and so forth. By the same token, if we make time slot $l$ a server time slot, then time slot $(l + \beta )\bmod N$ must be a server time slot, and so on and so forth. 
	
	\paragraph*{Case 2} If the period $k^* < N$ and $k^*$ is odd, then optimal packing is not possible. Without loss of generality, suppose that we make time slot $l$ a client time slot. Then time slot $(l + \beta )\bmod N$ must be a server time slot, time slot $(l + 2\beta )\bmod N$ must be a client time slot, and so on and so forth. We will reach a conclusion that time slot $l$ must also be a server time slot pairing with the client time slot $N - \beta $, leading to a conflict. Fig. \ref{fig_6} (c) shows the case of $\beta  = 2,N = 10$. The subring induced by $l = 0$ is $(0,2,4,6,8)$. If we make time slot $0$ a client time slot, then time slot $8$ must be client time slot, which implies that time slot $0$ must also be a server time slot, leading to a conflict. 
	
	\noindent 
	\paragraph*{Case 3} If the period $k^* < N$ and $k^*$ is even, then optimal packing is possible. There must be an index $v$ not on the subring induced by index $l$. This index $v$ can induce another subring with $k^*$ indexes that are distinct from the $k^*$ indexes of the subring induced by index $l$. By isomorphism, the period of the subring induced by $v$ must exactly be the same as the period of the subring induced by $l$ (i.e., on the ring there is no distinct difference between indexes as far as the situation that they are facing is concerned). The reason that these two subrings do not overlap is also obvious. If they have any overlapping index, then these two subrings will merge together to form a larger ring, contradicting the statement that index $v$ is not on the subring induced by index $l$. 
	
	In summary, optimal packing is possible if and only if $k^*$ is even as concluded in Proposition 1. As long as the current subrings do not cover all the indexes ${0, 1, ..., N-1}$, we can discover a new subring with $k^*$ indexes by inducing a subring from an indexes not yet in the existing subrings. Thus, in general, if $k^* < N$, then there must be $h$ subrings, each with $k^*$ distinct indexes, where 
	\begin{equation}
		\label{eqn:28}
		hk^* = N.
	\end{equation}
	Proposition 2 states that when optimal packing is possible, the $h$ subrings can be induced by indexes $0, 1, 2, ..., h-1$. To construct an optimal packing, we first list all the $h$ subrings. Then, on each subring, there are two possible packings. We can either let the subring-inducing index be a client time slot or a server time slot. Fig. \ref{fig_6} (d) shows the five subrings for the case of $\beta  = 5,N = 10$. The five subrings are (0,5), (1,6), (2,7), (3, 8), (4, 9). Together, they cover all the indexes in the set ${0, 1, ..., 9}$. 
	
	\begin{proposition}
		Let $0 < k^* \le N$ be the smallest integer such that 
		\begin{equation}
			\label{eqn:29}
			k^*\beta \bmod N = 0.
		\end{equation}
		Optimal packing is possible if and only if $k^*$ is even. 
	\end{proposition}
	\begin{IEEEproof}
		Obvious from the description of the above construction method. The $k^*$ here is basically the period of the subring mentioned in the construction method.
	\end{IEEEproof}
	
	\begin{proposition}
		When optimal packing is possible, suppose that for an $N$ and a $\beta $, we have $h$ distinct subrings, each with $k^*$ elements, $hk^*=N$. The $h$ distinct subrings can be found by inducing on index $0$, index $1$, ..., and index $h-1$. 
	\end{proposition}
	\begin{IEEEproof}
		Consider two subrings induced by two arbitrarily chosen indexes $e$ and $f$, $e,f \in \{ 0,1,...,h - 1\} $,$e \ne f$. If these two subrings were the same subring, then there would be two integers ${I_e}$ and ${I_f}$, ${I_e},{I_f} \in \{ 0,1,...,k^* - 1\} $, such that
		\begin{equation}
			\label{eqn:30}
			\begin{array}{l}
				(e + {I_e}\beta )\bmod N = (f + {I_f}\beta )\bmod N\\
				\Rightarrow \tau  + {g_1}N = ({I_e} - {I_f})\beta, 
			\end{array}
		\end{equation}
		for some integer $g_1$, where $\tau  = f - e \ne 0$. Furthermore, substituting \eqref{eqn:28} into \eqref{eqn:29}, we have 
		\begin{equation}
			\label{eqn:31}
			\begin{array}{l}
				k^*\beta \bmod kh = 0\\
				\Rightarrow \beta  = {g_2}h,
			\end{array}
		\end{equation}
		for some integer $g_2$. Substituting \eqref{eqn:28} and \eqref{eqn:31} into \eqref{eqn:30}, we then have
		\begin{equation}
			\label{eqn:32}
			\begin{array}{l}
				\tau  + {g_1}kh = ({I_e} - {I_f}){g_2}h,\\
				\tau  = (({I_e} - {I_f}){g_2} - {g_1}k^*)h.
			\end{array}
		\end{equation}
		Given that $1 \le \tau  < h$, it is not possible for \eqref{eqn:32} to hold.
	\end{IEEEproof}
	
	\subsubsection{What $N$ for a TDMA system is the most flexible in terms of optimal packing?} 
	
	An interesting question arises from the above discussion. Suppose that we want optimal packing to be possible for all $\beta  < N$. What $N$ will allow this?
	
	\begin{corollary}
		If $N$ is a power of $2$ (i.e., $N = {2^n}$ for some $n$), then optimal packing is possible for all $\beta  < N$. Conversely, if $N$ is not a power of $2$, then optimal packing is not possible for some $\beta  < N$. 
	\end{corollary}
	\begin{IEEEproof}
		We first prove the “if” part. First, suppose that $\beta $ is odd. Then there is no common factor between $\beta $ and ${N=2^n}$. The smallest $k^*$ that can satisfy the condition in proposition 1 must be $k^* = N$, which is even. Therefore, all odd $\beta $ allows for optimal packing. Next, suppose that $\beta $ is even. We can write $\beta  = \alpha {2^g}$ for some integer $g<n$ and for some odd integer $\alpha $. Thus, the common factor between $\beta$ and $N$ is $2^g$. The smallest $k^*$ such that (33) is satisfied must be $k^*=2^{n-g}$. Since $k^*$ is even, the condition in Proposition 1 is also satisfied. 
		
		We next prove the “only if” part (the converse). Since $N$ is even, if $N$ is not a power of $2$, we can express it as $N = m{2^g}$, where $m$ is an odd integer larger than $1$, and $g \ge 1$. Consider $\beta  = {2^g}$. In this case, the smallest $k^*$ that satisfies (33) is  $k^* = m$, which is odd. Thus, by Proposition 1, optimal packing is not possible with this $\beta$ (note: in general, there could be other $\beta$ for which optimal packing is not possible; this is just an example).
	\end{IEEEproof}
	
	\subsubsection{Discussion} 
	
	If $N$ is large and one client-server pair can have only one pair of time slots within the very large TDMA frame, that will be a serious limitation. In this case, we can always allocate several time-slot pairs to a client-server pair. For example, say $N = 64$ and $\beta  = 8$. If a client wants to interact with the server two times within a TDMA frame, we could assign two client-server time-slot pairs to the same client, say $(0,8)$ and $(32, 40)$. 
	
	\section{JIT System Implementation}	\label{sec:implementation}
	Our implementation\footnote{{https://github.com/Leo-Cheung-CUHK/openwifi-JIT}} is based on a modification of Openwifi\cite{jiao2020openwifi}, a free open-source IEEE 802.11 (Wi-Fi) SDR implementation on SoC platforms. Openwifi works on the Xilinx Zynq-7000 SoC that consists of a Field Programmable Gate Array (FPGA) and an ARM processor. Openwifi implements PHY and MAC functions with stringent latency requirements on FPGA\cite{jiao2020openwifi}. The Openwifi driver for accessing these PHY and MAC functions, on the other hand, is implemented in the embedded Linux running on the ARM processor. The Openwifi driver sends packets to and receives packets from the MAC layer using the Advanced eXtensible Interface 4 (AXI-4), a parallel high-performance low-latency SoC software-hardware communication interface. As with most Wi-Fi software implementations, the Openwifi driver instantiates Application Programming Interfaces (APIs) defined by the Linux mac80211 subsystem. Thanks to the modular design of Openwifi, researchers can study and modify this full-stack Wi-Fi implementation easily. Our modifications of Openwifi are described as follows
	
	To realize the JIT-triggered packet generation, we implemented the JIT middleware \textit{at the client} into the Openwifi driver in the ARM processor with four realized functionalities as discussed in the following:
	\begin{enumerate}[(i)]
		\item \textbf{Registering application}: An application running in the userspace first needs to register with the JIT middleware so that the JIT middleware can send pull signals to the application. Specifically, a client application should register its unique process ID (PID) with the JIT middleware. Since the JIT middleware is implemented within the Openwifi driver running in the kernel, the client application uses ioctl to register its PID with the JIT middleware (see https://en.wikipedia.org/wiki/Ioctl for details on ioctl). 
		\item \textbf{Scheduling pull-signal events}: The JIT middleware uses a Linux's high-resolution timer (hrtimer) with nanosecond resolution to schedule pull-signal events. The JIT system uses a reserved POSIX real-time signal as the pull signal. The reserved POSIX real-time signals can be the SIGRTMIN through SIGRTMAX signals intended for use defined by the users. For each pull-signal event, the JIT middleware sends a pull signal to the client's application using the SIGNAL Linux library.
		\item \textbf{Reacting to interrupt}: The client's application has a handler function. This handler function allows the client's application to generate and send a new packet to the TDMA MAC layer (see Fig. \ref{fig_5}) using the Openwifi driver API upon detecting the POSIX real-time signal.  
		\item \textbf{Returning feedback information}: Recall from the discussion in Section \ref{sec:JIT-triggered packet generation} that the MAC layer needs to return a calculated slack time to the JIT middleware. Thanks to the transmission-interrupt mechanism implemented by Openwifi in which the Openwifi driver always receives an interrupt after the TDMA MAC layer finishes a packet transmission. We use this interrupt mechanism to send the calculated slack time to the JIT middleware. 
	\end{enumerate} 
	Note that the server does not need JIT middleware since the JIT-triggered packet generation mechanism is to reduce the extra client's waiting time (see Section \ref{sec:motivation}). 
	
	We implemented a time-slotted TDMA MAC protocol on FPGA by overwriting the CSMA MAC protocol of Openwifi. In addition, we implemented a three-way handshake scheme based on precise time protocol (PTP) time to let each client-server pair have a common sense of time\footnote{The PTP algorithm of the scheme is the same as that of \cite{liang2021design}. Time synchronization error of our JIT system is in the order of inter IQ-sample time.}. 
	
	\section{Experimental Results} \label{sec:Experiment}
	As related in Section \ref{sec:motivation}, the JIT system minimizes the application-to-application RTT for client-server applications. This is done by minimizing: i) server's extra waiting time ${W_s}$ using the JIT time-slot allocation; ii) the client's extra waiting time ${W_c}$ using the JIT-triggered packet generation. In the following, we perform experiments to evaluate our JIT system. We detail the experiment set-up in Section \ref{sec:Experiment:1}, present the JIT time allocation in Section \ref{sec:Experiment:2}, validate the JIT-triggered packet generation in Section \ref{sec:Experiment:3}, and present the application-to-application RTT results in Section \ref{sec:Experiment:4}.

	\subsection{Experiment set-up} \label{sec:Experiment:1}
\begin{figure}[!htbp]
		\centering
		\includegraphics[width=3in]{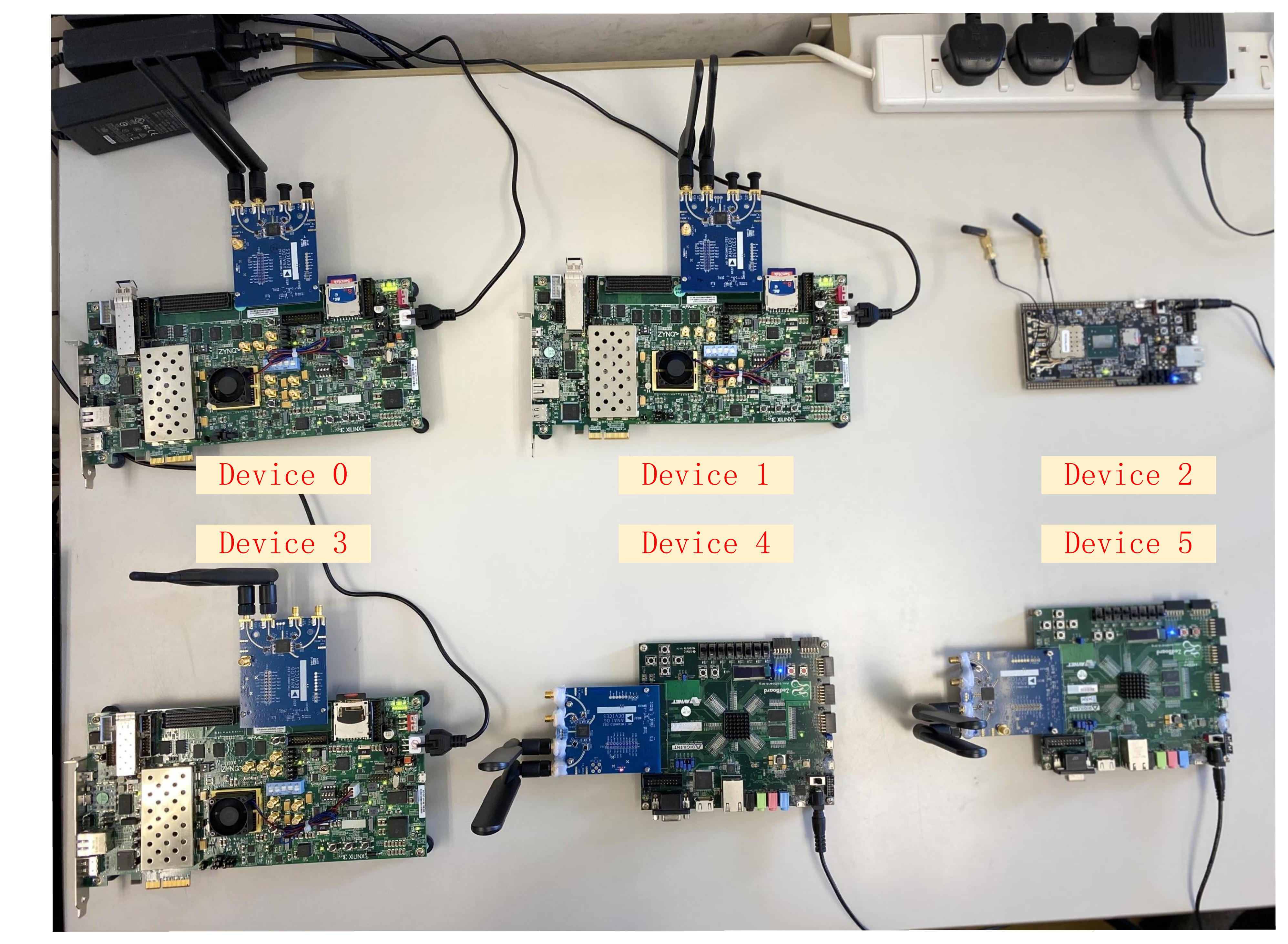}
		\caption{The experimental set-up.}
		\label{fig_7}
	\end{figure}

	As shown in Fig. \ref{fig_7}, six ZYNQ boards (consisting of three Xilinx ZC706 dev boards, one ADRV9361-Z7035 board and two Xilinx zed boards\cite{zynq}) were deployed as five client-server pairs. For Xilinx ZC706 dev boards and Xilinx zed boards, the AD-9361 RF evaluation boards of Analog Device Inc. were used to support wireless communication. In our experiment, Device $0$ in Fig. \ref{fig_7} is the server, and the rest of the devices are clients. 
	
	Several settings that apply to this experiments are as follows: i) the payload size of request and response are the same; ii) the time duration of a time-slot is set to the same as that of request and response; iii) the server and clients make use of a non-real-time Linux OS. Other details related to the values of the system parameters are given in Table \ref{tab:1}.
	
	\begin{table}[]
		\caption{PARAMETERS OF THE EXPERIMENT}
		\label{tab:1}
		\begin{tabular}{|l|l|}
			\hline
			Payload Size (of request and response)                    & $126$ Bytes               \\ \hline
			Packet Duration/ Time Duration of a Time Slot ($\Delta t$)& $150 \mu {\rm{s}}$ \\ \hline
			Maximum preemption delay (${ST}_{\rm{target}}$)\tablefootnote{Before this experiment, we pre-estimated the maximum preemption delay ${ST}_{\rm{target}}$ on the client as discussed in Section \ref{sec:JIT-triggered packet generation}, and we set $S{T_{{\rm{target}}}} = 30 \mu {\rm{s}}$.} 		
			&$30\mu {\rm{s}}$   \\ \hline
			Time for request/response generation ($D_c$ and $D_s$)    &$30 \mu {\rm{s}}$   \\ \hline
			Number of Slots per TDMA Frame ($N$)					  &$64$                       \\ \hline
			Time Duration of a TDMA Frame						      &$9.6 {\rm{ms}}$            \\ \hline
			{Constant Smoothing Parameter $\alpha $}  &{$0.9$}     \\ \hline
			{Initial Smoothed Statistic ${\hat n_0}$} &{$0 \mu {\rm{s}}$}            \\ \hline
			{Number of Packets Used During Initialization  $Q$} &{$400$}            \\ \hline	
		\end{tabular}
	\end{table}
	
	\subsection{Deployment of JIT time-slot allocation mechanism} \label{sec:Experiment:2}
	The time slots preallocated to the five client-server pairs are determined as follows:
	\begin{enumerate}[(i)] 
		\item \textbf{Obtain server processing delay ${D_s}$:} Given the above time slot duration $\Delta t$ of our system, we find that our platform is such that the server processing delay is within one time slot, i.e., $0 < {D_s} < \Delta t$. 
		\item \textbf{Derive inter time-slot parameter $\beta$:} With $0 < {D_s} < \Delta t$, we derive that $\beta  = 2$ from \eqref{eqn:25}.
		\item \textbf{Allocate time slots:} In this setup, we have $\beta = 2$ and $N = 64$. According to \eqref{eqn:29} in Proposition 1, optimal packing is possible if and only if $k^{*} = 32$. Using \eqref{eqn:28}, we derive that there must be two subrings in this optimal packing. According to Proposition 2, there are 32 possible time-slot pairs. And we used five out of them for our experiment. That is, the time-slot pairs allocated to the five client-server pairs are $(0,2)$, $(1,3)$, $(4,6)$, $(5,7)$, and $(8, 10)$.
	\end{enumerate} 
	
	\subsection{Validation of JIT-triggered packet generation mechanism} \label{sec:Experiment:3}
	This subsection investigates whether the JIT-triggered packet generation can minimize ${W_c}$ in the face of the clock offset and timing imprecision. We benchmark our JIT system against a baseline system without the JIT-triggered packet generation mechanism. We describe the similarities and differences between the JIT system and the baseline system below:
	
	\noindent
	\textbf{\textit{Similarities}}
	\begin{itemize}
		\item The client's TDMA MAC layer schedules one request transmission per TDMA frame using the MAC clock. 
		\item Upon receiving a request, the server generates a response.
		\item The server's TDMA MAC layer schedules one response transmission per TDMA frame using the MAC clock. 
	\end{itemize} 
	
	\noindent
	\textbf{\textit{Differences}}
	\begin{itemize}
		\item In the JIT system, the client's JIT middleware schedules one pull-signal event per TDMA frame using the JIT clock. A request is generated by the client's application if and only if the client's application is interrupted by the pull signal. 
		\item In the baseline system, there is no pull signal. The client's application itself schedules one request generation per TDMA frame using the JIT clock.
	\end{itemize} 
	
	For a comprehensive investigation on the clock-offset issue, we compare the JIT system and baseline system under three different \textit{controlled} settings on the clock tick rates: i) in \textbf{setting 1}, the JIT clock and the MAC clock have the same tick rate; ii) in \textbf{setting 2}, the JIT clock ticks more slowly than the MAC clock; iii) in \textbf{setting 3}, the JIT clock ticks faster than the MAC clock. For setting 2 (setting 3), we assume that the clock tick rate difference in the two clock systems is a small constant value $+0.0005\%$ ($-0.0005\%$). Most oscillators, including low-cost oscillators, have frequency accuracy better than this specification. 
	
	In practice, we cannot directly control the relative tick rates of the JIT clock and MAC clock in a general framework (see Section \ref{sec:JIT-triggered packet generation}). However, our JIT system design is based on the Openwifi design in which the same clock source is used to drive the Linux OS and the communication layer, and there will be no asynchrony between the two clocks\footnote{If we deploy an embedded system design like Openwifi, we do not need to address the clock-offset issue discussed in Section \ref{sec:JIT-triggered packet generation}. However, we still need the JIT middleware to address the time-imprecision issue due to preemption delays described in Section \ref{sec:JIT-triggered packet generation}. As we will see in Section \ref{sec:Experiment:3} and Section \ref{sec:Experiment:4}, the system could still suffer from large extra client's waiting-times if the time-imprecision issue is not addressed.}. Thus, for the study of the general framework, we simulated the settings in an artificial way. First, setting 1 requires no further effort since the ZYNQ boards are SoC evaluation boards in which the JIT clock and MAC clock share the same clock oscillator and tick at the same rate. To simulate the clock tick rate differences in both settings 2 and 3, we performed some adjustments as follows: i) in the JIT system, the initial TDMA frame duration of the client's JIT middleware was adjusted to $9.6048{\rm{ms}}$ and $9.5952{\rm{ms}}$ to simulate $ - 0.0005\% $ and $0.0005\% $ clock tick rate differences in the two clock systems, respectively; ii) in the baseline system, the TDMA frame duration of the client's application was adjusted to $9.6048{\rm{ms}}$ and $9.5952{\rm{ms}}$ for settings 2 and 3, respectively. 
	
	We expect the JIT-triggered packet generation of the JIT system can still make sure that, on average, the pull-signal period is still $9.6\rm{ms}$ for both settings 2 and 3. For clarity, we use the notation $\rm{JIT}_a$ and the notation $\rm{Baseline}_a$ to denote the JIT system and the baseline system in the setting $a$, respectively. Meanwhile, since a non-real-time OS was deployed, both the JIT system and baseline system for all three settings could experience preemptions from other tasks. Hence, this experiment could also investigate the effect arising from the timing-imprecision issue. Without loss of generality, we use pair $1$ with time-slot pair $(0, 2)$ for the investigation. 
	
\begin{figure}[!htbp]
		\centering
		\includegraphics[width=3.5in]{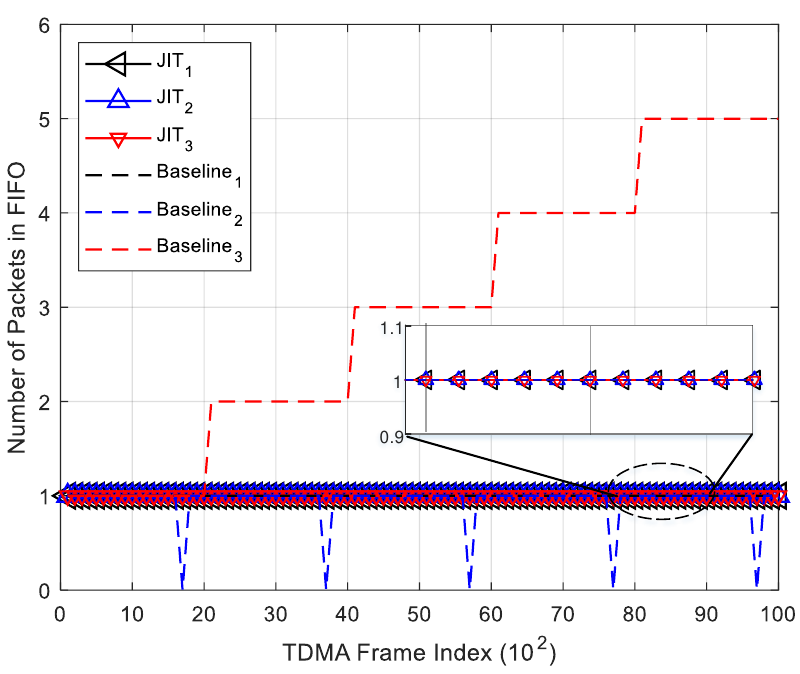}
		\caption{{The number of packets in the transmission FIFO buffer collected once every $100$ TDMA frames. Note that the points for $\rm{JIT}_1$, $\rm{JIT}_2$, $\rm{JIT}_3$, and $\rm{Baseline}_1$ overlap – the buffer occupancy stays constant at $1$ throughout.}}
		\label{fig_8}
	\end{figure}

	In a general full-stack Wi-Fi system, a packet generated by an upper layer will first be stored in a FIFO buffer at a lower layer, waiting to be retrieved by a process at the lower layer for transmission. For example, in the design of Openwifi, a packet sent by the ARM first arrives at a transmission FIFO buffer in the FPGA. This packet is later retrieved by the MAC layer for RF transmission. Fig. \ref{fig_8} plots the number of packets in the transmission FIFO buffer for the JIT system and baseline system (the y-axes) when the time slot allocated to the client comes along (the x-axes). To avoid cluttering, Fig. \ref{fig_8} only plots the buffer occupancy once every $100$ TDMA frames. We can see from Fig. \ref{fig_8} that
	\begin{enumerate}[(i)]
		\item Both $\rm{Baseline}_1$ and $\rm{JIT}_1$ hold one packet in the transmission FIFO buffer all the time. 
		\item In $\rm{Baseline}_2$, the number of packets in the transmission FIFO buffer wraps from $0$ to $1$ repeatedly. The reason of the wrapping is that the FIFO buffer may underflow in $\rm{Baseline}_2$ once in a while: when the transmission opportunity comes along at the TDMA MAC layer, there is no packet in the FIFO buffer. In contrast, $\rm{JIT}_2$ holds one packet all the time.
		\item In $\rm{Baseline}_3$, the number of packets in the transmission FIFO buffer keeps increasing. The FIFO buffer overflows in $\rm{Baseline}_3$. However, $\rm{JIT}_3$ still holds one packet all the time.
	\end{enumerate} 
	The ${W_c}$ in $\rm{Baseline}_3$ will eventually go to infinity, leading to an unbounded application-to-application RTT. In particular, $\rm{Baseline}_3$ could crash due to the FIFO buffer overflow. Therefore, we bypass the experimental results of $\rm{Baseline}_3$ going forward. 
	
\begin{figure}[!htbp]
		\centering     
		\includegraphics[width=3.5in]{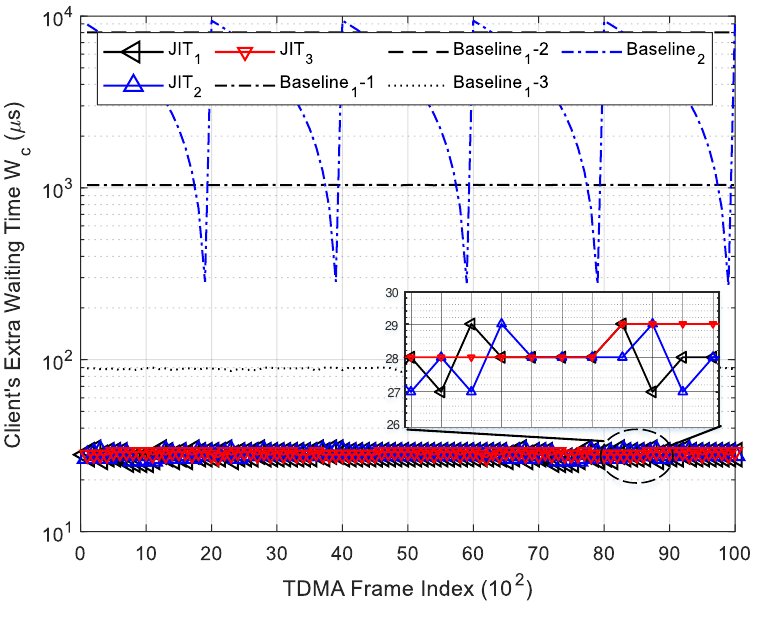}
		\caption{{Plots of the client's extra waiting time ${W_c}$ collected once every $100$ TDMA frames}}
		\label{fig_9}
	\end{figure}

	We next show the ${W_c}$ of the JIT system and baseline system (except $\rm{JIT}_3$) in Fig. \ref{fig_9}. In Fig. \ref{fig_9}, we only plot the ${W_c}$ once for every $100$ TDMA frames. From Fig. \ref{fig_9}, we have the following observations:
	\begin{enumerate}[(i)]
		\item The ${W_c}$ of $\rm{JIT}_1$, $\rm{JIT}_2$ and $\rm{JIT}_3$ are stable and roughly the same. Specifically, we can see that despite the fluctuations of processing delays due to preemptions, the ${W_c}$ of the JIT system only fluctuates within a small range around $30\mu \rm{s}$ (the value of ${ST}_{\rm{target}}$). This result is consistent with the discussion in Section \ref{sec:JIT-triggered packet generation}. The experimental results confirm that the JIT system is robust against preemption jitters: it effectively addresses the timing-imprecision issue and keeps ${W_c}$ stable at the target slack time ${ST}_{\rm{target}}$
		\item The ${W_c}$ of $\rm{Baseline}_2$ wraps from the minimum to the maximum repeatedly. A packet may arrive at the FIFO buffer just slightly after the last transmission opportunity. In particular, the ${W_c}$ in $\rm{Baseline}_2$ ranges from $0$ to the value of one TDMA frame duration.
		\item $\rm{Baseline}_1$, without the clock-tick rate difference, has a more stable ${W_c}$ compared with that of $\rm{Baseline}_2$. However, unlike the other systems, Baseline systems or JIT systems, the ${W_c}$ obtained in each experimental trial of $\rm{Baseline}_1$ can be different and subject to randomness. For example, Fig. \ref{fig_9} shows the results of three experiments for $\rm{Baseline}_1$. These observed results are explained as follows. For $\rm{Baseline}_1$, the moment the application layer generates a packet is not synchronized to the moment of the next transmission opportunity. Within the same experiment, the offset in packet-generation time and transmission opportunity is constant throughout since the application layer and the MAC layer use the same clock. However, the offset itself is random from experiment to experiment. Thus, depending on this random offset, the ${W_c}$ experienced is also random and ranges from zero to one TDMA frame duration. Overall, the average value of $\rm{Baseline}_1$'s ${W_c}$, which is half of a TDMA frame duration, is much larger than the stable ${W_c}$ of $\rm{JIT}_1$. 
	\end{enumerate} 
	This result convinces us that the JIT-triggered packet generation mechanism maintains good synchronization between the JIT middleware and the TDMA MAC layer, addressing the clock-offset issue and ensuring each packet can be transmitted at the desired time.
	
	\subsection{Application-to-Application RTT Evaluation} \label{sec:Experiment:4}
	This subsection first investigates how ${W_c}$ affects the application-to-application RTT of the JIT system and the baseline system. To obtain the application-to-application RTT, we logged two kinds of timestamps on the client's Linux system: the time when the client's application starts to generate a request packet, denoted by $T^{start}$, and the time when the client's application receives the corresponding response packet, denoted by $T^{end}$. The application-to-application RTT for a client-server application is $RTT = T^{end} - T^{start}$.  
	
\begin{figure}[!t]
		\centering
		\includegraphics[width=3.5in]{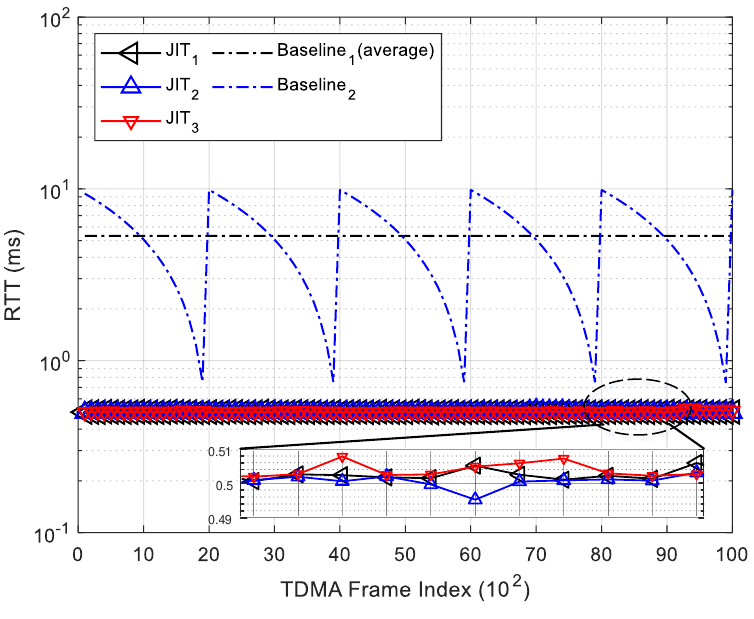}
		\caption{{$RTT$ of the JIT system and the baseline system collected once every $100$ TDMA frames.}}
		\label{fig_10}
	\end{figure}

	In this experiment, both the baseline system and JIT system use JIT time-slot allocation. And we again use client-server pair 1 assigned with the time-slot pair $(0, 2)$ for the investigation. Fig. \ref{fig_10} plots the $RTT$ once every $100$ TDMA frames. We have the following observations: 
	\begin{enumerate}[(i)]
		\item Having a stable ${W_c}$, $\rm{JIT}_1$, $\rm{JIT}_2$ and $\rm{JIT}_3$ all have small and stable $RTT$s. 
		\item The $RTT$ of $\rm{Baseline}_2$ drifts with time due to the varying ${W_c}$. In particular, the maximum value of the $RTT$ in $\rm{Baseline}_2$ could be more than one TDMA frame duration. 
		\item Recall that the ${W_c}$ of $\rm{Baseline}_1$ can change from experiment to experiment. Thus, the $RTT$ can also change from experiment to experiment. In Fig. \ref{fig_10}, we plot the average $RTT$ of $\rm{Baseline}_1$ among these experiments. As shown, the average $RTT$ of $\rm{Baseline}_1$ is much larger than the stable $RTT$ of $\rm{JIT}_1$. 
	\end{enumerate} 
	
	Note that even with the use of JIT time-slot allocation here, the $RTT$s of all the three baseline systems are still larger than that of the JIT systems. If the baseline systems did not use the JIT-time slot allocation, then their $RTT$s would be even larger due to the effect of ${W_s}$. For example, if we assign the time-slot pair $(0, 32)$ rather than $(0, 2)$ to the two baseline systems, then their $RTT$ will increase by the duration of 30 TDMA slots, which is $4.5 ms$ in our system. If we are unfortunate enough to assign the time-slot pair$ (0, 1)$ to the two baseline systems, then their $RTT$ will increase by $62$ TDMA slots (around one TDMA frame) which is $9.3 ms$ in our system. This case corresponds to the case in which the conventional networking system has the worst-case maximum possible server's waiting time ${W_s}$, as shown in \eqref{eqn:8} of Section \ref{sec:motivation}. In short, the worst-case $RTT$ of baseline systems depends on the TDMA frame duration. The larger the TDMA frame duration a baseline system has, the larger the worst-case $RTT$. However, this situation does not happen in our JIT system. That is, the $RTT$ of our JIT system is independent of the TDMA frame duration. As seen in Fig. \ref{fig_10}, the $RTT$ of the JIT system is stable at around $510\mu s$. 
	
	We further decompose this $510\mu s$ of $RTT$ into several components as listed in the following:
	\begin{enumerate}[(i)]
		\item $30\mu s$ for the client to generate a request packet, as given in Table \ref{tab:1}.
		\item $30\mu s$ of the target slack time ${ST}_{\rm{target}}$ for each request packet. Note that the value of ${ST}_{\rm{target}}$ depends on the maximum preemption delay that a system could experience.
		\item $450\mu \rm{s}$ of the elapsed time between a request packet transmitted by the client at the beginning of time slot $0$ and the corresponding response packet received by the client at around the end of time slot $2$.
		\item Some other minor time that the client took to return the response back to the application.
	\end{enumerate} 
	The above discussions indicate that the JIT system as we implemented has achieved the minimum possible $RTT$.
	
	In summary, the JIT system enables synchronized JIT-triggered packet generation with which a packet can be sent shortly after being generated by the application. Together with an optimal time-slot allocation derived from our JIT time-slot allocation scheme, our JIT system pushes the application-to-application RTT of a client-server application to the limit. 
	
	\section{Conclusion} \label{sec:conclusion}
	Due to loose coordination between the communication and application layers, client-server applications running on conventional networking systems could suffer from large request-response latency. This round-trip latency includes the extra waiting time incurred by the request while it waits for a transmission opportunity at the client side, and the extra waiting incurred by the response while it waits for a transmission opportunity at the server side. What is even worse for the conventional networking systems is that, in a TDMA system, these two extra waiting times depend on the duration of the TDMA frame: the larger the TDMA frame, the larger the worst-case request-response latency the client-server application can suffer from. 
	
	To reduce request-response latency, we put forth a JIT framework with two mechanisms to minimize the two extra waiting times: i) a JIT-triggered packet generation at the client side to inform the client application of an upcoming transmission opportunity so that the client application can generate a just-in-time request; ii) a JIT time-slot allocation that caters to the delay in between the reception of the request and the arrival of the response at the communication layer at the server, so that when the response is generated by the server application, there is a just-in-time transmission opportunity to send out the response. Our mathematical analysis indicates that: i) the total extra waiting times in conventional networking systems can be as large as two times the duration of the TDMA frame; ii) the JIT-triggered packet generation mechanism is robust against the preemption delay in multitasking software architecture and against the timing offset among different system components; iii) a TDMA network with a power-of-2 time slots per superframe is optimal for realizing the JIT time-slot allocation. 
	
	We implemented our JIT system on the Xilinx SoC platform. The experimental results validate our analysis, and show that: i) the total extra waiting times in our JIT framework do not depend on the duration of the TDMA frame and only fluctuate within a small range; ii) the JIT system can achieve the minimum achievable application-to-application RTT. We believe that our JIT framework provides a useful paradigm for many time-sensitive industrial network applications with very stringent latency requirements for client-server interactions.
	
	{We remark that the JIT framework can potentially be incorporated into next-generation networks, e.g., the 5G network, to support user applications that require low latency \cite{hu2017iot,liu2020task,minoli2017iot}. Regarding 5G, the JIT-triggered packet-generation mechanism can be leveraged to decrease user application-layer latency. To this end, once the PHY-layer resource allocation to a specific user node is determined (e.g., after fixing the transmit time interval (TTI) \cite{marsch20185g} of the 5G network \cite{elsayed2019ai}), the user node’s MAC layer will return the packet waiting times (the calculated slack times as discussed in Section V) at the MAC layer to the JIT middleware recurrently. JIT middleware could then inform the application to generate packets in a JIT manner so as to reduce the MAC-layer waiting time.}
	
	{Finally, although this paper focuses on the client-server messaging model, the JIT system could also run in other common messaging models. For example, in the publish-subscribe model\footnote{{https://en.wikipedia.org/wiki/Publish\%E2\%80\%93subscribe\_pattern}}, a group of sensors may send information to a broker to be distributed to subscribers. If the sensors make their measurements in a JIT manner (i.e., just before their transmission opportunities arise) and send them to the broker, then the broker can disseminate the freshest possible measurement information to the subscribers.}
	
	\appendices
	\section{Proof of clock-tick difference compensation and stability} \label{appendix:I}
	In the following, we show that the adaptive synchronization mechanism in our JIT-triggered packet generation can compensate for this clock-tick difference and is stable. We first modify \eqref{eqn:21} to a more convenient form in \eqref{eqn:33} for later analytical development. Specifically, substituting \eqref{eqn:18} into \eqref{eqn:21}, we have 
	\begin{equation}
		\label{eqn:33}
		U_i^{JIT} = U_{i - 1}^{JIT} + ({o_{i - 1}} - {o_i}) + F + {\hat n_i}, i \ge 1.
	\end{equation}
	From \eqref{eqn:20}, we have
	\begin{equation}
		\label{eqn:34}
		{ST}_i = U_i^{MAC} - (U_i^{JIT} + D_{c,i}).
	\end{equation}
	Substituting \eqref{eqn:33} and \eqref{eqn:15} into \eqref{eqn:34}, we have
	\begin{align}
		{ST}_i &= (U_{i - 1}^{MAC} - U_{i - 1}^{JIT}) - (({o_{i - 1}} - {o_i}) + {{\hat n}_i} + D_{c,i})\\
		&= (ST_{i - 1} + D_{c,i - 1}) - (({o_{i - 1}} - {o_i}) + {{\hat n}_i} + D_{c,i})\\
		&= ST_{i - 1} + \Delta {o_i} - {{\hat n}_i} - \Delta {D_{c,i}}, i \ge 1 \label{eqn:35}
	\end{align}
	where $\Delta {o_i} = {o_i} - {o_{i - 1}}$ and $\Delta {D_{c,i}} = D_{c,i} - D_{c,i - 1}$. From \eqref{eqn:24}, we also have 
	\begin{equation}
		\label{eqn:36}
		{\hat n_i} = \left\{ {\begin{array}{*{20}{c}}
				{\alpha( S{T_{i - 1}} - \hat{ST}_{\rm{target}}  )}&{n = 1}\\
				\begin{array}{l}
					\alpha(S{T_{i - 1}} - \hat{ST}_{\rm{target}}) + \\
					\alpha(1 - \alpha)(S{T_{i - 2}} - \hat{ST}_{\rm{target}}) + ...\\
					+ \alpha{(1 - \alpha)^{i - 1}}(S{T_0} - \hat{ST}_{\rm{target}})
				\end{array}&{n > 1}
		\end{array}}. \right.
	\end{equation}

	Now, define the slack time deviation as $\tilde{ST}_i = {ST}_i - \hat{ST}_{\rm{target}}$. Substituting \eqref{eqn:36} into \eqref{eqn:35} and subtracting $\hat{ST}_{\rm{target}}$ for both sides of \eqref{eqn:35}, we have
	\begin{equation}
		\label{eqn:37}
		\begin{array}{l}
			{\tilde{ST}_i = {ST}_i - \hat{ST}_{\rm{target}}}\\
			= {ST}_{i - 1} - \hat{ST}_{\rm{target}} + \Delta {o_i} - \Delta {D_{c,i}} - {{\hat n}_i}\\
			{\rm{       =  }}\left\{ {\begin{array}{*{20}{c}}
					{\Delta {o_1} - \Delta {D_{c,1}} + (1 - \alpha)\tilde{ST}_0}&{n = 1}\\
					\begin{array}{l}
						\Delta {o_i} - \Delta {D_{c,i}} + (1 - \alpha)\tilde{ST}_{i-1}\\
						- \alpha(1 - \alpha)\tilde{ST}_{i-2} - \alpha{(1 - \alpha)^2}\tilde{ST}_{i-3}\\
						... - \alpha{(1 - \alpha)^{i - 1}} \tilde{ST}_0
					\end{array}&{n > 1}
			\end{array}}. \right.
		\end{array}
	\end{equation}
	In general, we see that the deviation $\tilde{ST}_i$ fluctuates with the instantaneous clock offset increment $\Delta {o_i}$ and the instantaneous preemption delay increment $\Delta {D_{c,i}}$. We next show that the system is stable. 
	
	First, define our system input as ${x_i} = \Delta {o_i} - \Delta {D_{c,i}}$ and our system output as ${y_i} = \tilde{ST}_i$. Then we can rewrite \eqref{eqn:37} in the form of difference equations as 
	\begin{equation}
		\label{eqn:38}
		{{\rm{y}}_i}{\rm{ = }}\left\{ {\begin{array}{*{20}{c}}
				{{x_1} + (1 - \alpha){{\rm{y}}_0}}&{i = 1}\\
				{{x_i} + (1 - \alpha){y_{i - 1}} - \alpha\sum\limits_{j = 0}^{i - 2} {{{(1 - \alpha)}^{i - j - 1}}{{\rm{y}}_j}} }&{i > 1}
		\end{array}}. \right.{\rm{ }}
	\end{equation}
	where ${{\rm{y}}_0} = {\rm{S\tilde T}} = C < \infty$ is the initial state of the system. From \eqref{eqn:38} we can see that this system is a discrete-time linear feedback system where the output is generated not only from the input but also from the previous outputs. From \eqref{eqn:38}, we can write the z-transform of the system response as 
	\begin{equation}
		\label{eqn:39}
		\begin{array}{l}
			H(z) = \frac{{Y(z)}}{{X(z)}} = \frac{1}{{1 - (1 - \alpha){z^{ - 1}} + \alpha(1 - \alpha){z^{ - 2}} + \alpha{{(1 - \alpha)}^2}{z^{ - 3}} + ...}}\\
			= \frac{1}{{1 - {z^{ - 1}} + \frac{{\alpha{z^{^{ - 1}}}}}{{1 - (1 - \alpha){z^{^{ - 1}}}}}}} = \frac{{{z^2} - (1 - \alpha)z}}{{{z^2} - 2(1 - \alpha)z + (1 - \alpha)}}
		\end{array}.
	\end{equation}
	From \eqref{eqn:39}, we can derive the Z-plane's poles ${r_1} = (1 - \alpha) + j\sqrt {\alpha(1 - a)} $ and ${r_2} = (1 - \alpha) - j\sqrt {\alpha(1 - \alpha)} $. Since the input signal of our system ${x_i}$ is a right-sided sequence and is rational, the region of convergence (ROC) of our system is the region in the z-plane outside the outermost pole. Also, since $\left| {{r_1}} \right| = \left| {{r_2}} \right| < 1$, our system's ROC includes the unit circle. According to the stability sufficient condition, our system is hence stable. Furthermore, if ${x_i} = \Delta {o_i} - \Delta {D_{c,i}} = \Delta$, where $\Delta$ is a constant value, ${ST}_i$ will finally converge to  $\Delta  + \hat{ST}_{\rm{target}}$. 
	
	\section{Proof of bounded differential $U_i^{MAC} - U_i^{JIT}$}	\label{appendix:II}
	In the following, we show that the adaptive synchronization mechanism in our JIT-triggered packet generation can have bounded differential $U_i^{MAC} - U_i^{JIT}$ even if the JIT clock and the MAC clock are not synchronized and can drift apart. 
	
	Recall that $U_i^{JIT}$ is the time when the system generates a pull signal for requesting the information, $U_i^{MAC}$ is the time when the system transmits the information. When the system reaches convergence where ${ST}_i = \Delta  + \hat{ST}_{\rm{target}}$, from \eqref{eqn:34} we have $U_i^{MAC} - U_i^{JIT} = \Delta  + \hat{ST}_{\rm{target}} + D_{c,i} = \Delta {o_i} + D_{c,i - 1} + \hat{ST}_{\rm{target}}$. If $\Delta {o_i} - \Delta {D_{c,i}} = \Delta$ is constant, we have two particular extreme cases: 
	\begin{enumerate}[(i)] 
		\item If $\Delta {D_{c,i}} = 0$ and $\Delta {o_i} = \Delta$, the system has no jitter in preemption delay and $U_i^{MAC} - U_i^{JIT}$ is constant.
		\item If $\Delta {D_{c,i}} =  - \Delta$ and $\Delta {o_i} = 0$, $U_i^{MAC} - U_i^{JIT}$ drifts with $D_{c,i - 1}$. In particular, when $\Delta {D_{c,i}} =  - \Delta  \ge 0$, $U_i^{MAC} - U_i^{JIT}$ increases as the value of preemption delay increases. In this case, $U_i^{MAC} - U_i^{JIT}$ becomes larger and larger even though ${ST}_i$ converges. 
	\end{enumerate} 
	
	In general, if the preemptive delay is bounded, case (ii) is only a hypothetical scenario. The discussion of case (i) indicates that the system can have bounded differential $U_i^{MAC} - U_i^{JIT}$ even if the JIT clock and the MAC clock can drift apart. That is, our system can deal with the clock asynchronies between two different clock systems. 
	
	\section{JIT in CSMA networks}	\label{appendix:III}
	This appendix considers what if a CSMA network is used instead of a TDMA network. Instead of a deterministic JIT system, we can aim for a probabilistic JIT system, since the underlying network does not guarantee communication resources to the users. For CSMA networks such as Wi-Fi, the MAC layer has a back-off counter to regulate when a node gets a transmission opportunity. When the counter value reaches zero, then the node gets to transmit. For JIT principle A, when the counter reaches a value equivalent to the max turn-around time for the application to generate a packet and deliver it to the PHY layer, then the JIT middleware can send a pull signal to the application layer. This maneuver is probabilistic in that the counter may still get frozen because the node senses another node transmitting before the counter value reaches zero. In this case, the packet would have arrived at the PHY layer too early. However, we remark that using the probabilistic JIT pull is still better than not using it. If the application layer just pushes data to the MAC layer regardless of the counter value, and if a packet arrives at the MAC layer with a large counter value, the chances for the counter to be repeatedly frozen by a series of packets would be even higher, resulting in an even higher wait time at the MAC layer.
	
	For JIT principle B, a possibility is to set the counter value at the server to correspond to the maximum turn-around time for the server to process the request and generate a response to be delivered to the MAC layer. That is, upon receiving a response packet, the counter value is set to this value. Again, this is a best-attempt probabilistic maneuver.
	
	Yet another approach is to redesign the back-off mechanism of the CSMA network (assuming it only supports one server and multiple clients as in our use case here) so as to minimize the delay experienced by the client and the delay experienced by the server, taking into account the JIT mechanisms. This line of work awaits future investigation.
	
	In general, we believe that the JIT principles are applicable to all networks in which (i) the MAC layer of a client has a sense of when the next transmission opportunity may be upcoming (for JIT principle A), and (ii) the MAC layer of the server has a certain priority in scheduling the next transmission opportunity for itself (for JIT principle B). For example, a network that allocates resources based on polling or reservation can also adopt the JIT principles.

	
	\bibliographystyle{IEEEtran}
	\bibliography{database}

	\ifCLASSOPTIONcaptionsoff
	\newpage
	\fi

\end{document}